%% file: lmcs.tex
\documentclass{CSML}
\pdfoutput=1

\usepackage{lastpage}
\newcommand{\dOi}{14(1:8)2018}
\lmcsheading{}{1--\pageref{LastPage}}{}{}%
{Dec.~07, 2016}{Jan.~22, 2018}{}

\usepackage{xspace}
\usepackage{bm}
\usepackage{tikz}
\usepackage{amsmath}
\usepackage{amsthm}
\usepackage{amsfonts}
\usepackage{etoolbox}
\usepackage{hyperref}
\hypersetup{hidelinks}
\usepackage{url}

\definecolor{drawcolour}{rgb}{0.75,0.75,0.75}
\definecolor{fillcolour}{rgb}{1,1,1}

\tikzset{blob/.style={circle, draw, fill=white, inner sep=2pt}}
\newcommand\N{\ensuremath{\mathbb{N}}}
\tikzset{morphism/.style={draw, circle, fill=white, inner sep=0pt, minimum width=13pt, font=\scriptsize, draw=black, opacity=1}}
\tikzset{region/.style={draw=drawcolour, fill=fillcolour, thick, fill opacity=0.7}}

\newtheorem{theorem}{Theorem}

\theoremstyle{definition}
\newtheorem{definition}[theorem]{Definition}
\newtheorem{example}[theorem]{Example}

\newcommand\ignore[1]{}

\tikzstyle{dotpic}=[scale=0.4]
\tikzstyle{dot}=[circle,draw,inner sep=0pt,minimum width=2mm,thick]
\tikzstyle{white dot}=[dot, fill=white]
\tikzstyle{blue dot}=[dot, fill=blue]
\tikzstyle{red dot}=[dot, fill=red]
\tikzstyle{gray edge}=[gray]
\tikzstyle{none}=[]
\pgfdeclarelayer{background}
\pgfdeclarelayer{edgelayer}
\pgfdeclarelayer{nodelayer}
\pgfsetlayers{background,edgelayer,nodelayer,main}
\tikzstyle{none}=[inner sep=0mm]
\tikzstyle{every loop}=[]
\tikzstyle{mark coordinate}=[inner sep=0pt,outer sep=0pt,minimum size=3pt,fill=black,circle]

\newcommand{\dpair}[1]{{\color{blue!60!black}\left\langle \color{black} #1
  \color{blue!50!black} \right\rangle}}

\newcommand{\hrefLMCS}[2]{\href{#1}{\tt #2}}

\makeatletter
\def\calign@preamble{%
   &\hfil\strut@
    \setboxz@h{\@lign$\m@th\displaystyle{##}$}%
    \ifmeasuring@\savefieldlength@\fi
    \set@field
    \hfil
    \tabskip\alignsep@
}
\let\cmeasure@\measure@
\patchcmd\cmeasure@{\divide\@tempcntb\tw@}{}{}{}
\patchcmd\cmeasure@{\divide\@tempcntb\tw@}{}{}{}
\patchcmd\cmeasure@{\ifodd\maxfields@
  \global\advance\maxfields@\@ne
  \fi}{}{}{}    
\newenvironment{calign}
{%
  \let\align@preamble\calign@preamble
  \let\measure@\cmeasure@
  \align
}
{%
  \endalign
}  
\newenvironment{calign*}
{%
  \let\align@preamble\calign@preamble
  \let\measure@\cmeasure@
  \align
  \notag
}
{%
  \endalign
}  
\makeatother

\usepackage{microtype}

\newcommand{\globular}{\textsf{Globular}\xspace}

\def\Sig{\ensuremath{\mathrm{Sig}}}
\def\Set{\ensuremath{\mathrm{Set}}}

\def\Diag{\ensuremath{\mathrm{Diag}}}
\def\List{\ensuremath{\mathrm{List}}}
\def\I{\text{I}}
\def\II{\text{II}}
\def\III{\text{III}}
\def\IV{\text{IV}}
\def\V{\text{V}}
\def\VI{\text{VI}}

\begin{document}

\title{Globular: an online proof assistant for higher-dimensional rewriting
}
\author{Krzysztof Bar}
\address{Department of Computer Science, University of Oxford}
\email{krzysztof.bar@cs.ox.ac.uk}

\author{Aleks Kissinger}
\address{Institute for Computing and Information Sciences, Radboud University Nijmegen}
\email{aleks@cs.ru.nl}

\author{Jamie Vicary}
\address{Department of Computer Science, University of Oxford}
\email{jamie.vicary@cs.ox.ac.uk}
%
%
%
%
%
%
%

\maketitle

\begin{abstract}
This article introduces \globular, an online proof assistant for the formalization and verification of proofs in higher-dimensional category theory. The tool produces graphical visualizations of higher-dimensional proofs, assists in their construction with a point-and-click interface, and performs type checking to prevent incorrect rewrites. Hosted on the web, it has a low barrier to use, and allows hyperlinking of formalized proofs directly from research papers. It allows the formalization of proofs from logic, topology and algebra which are not formalizable by other methods, and we give several examples.
\end{abstract}

\section{Introduction}

This paper is a system description for \globular~\cite{globular}, an online tool for formalizing and verifying proofs in semistrict globular higher category theory. It operates from the perspective of \textit{higher-dimensional rewriting}, with terms represented as graphical structures, and proofs constructed and visualized as sequences of rewrites on these structures. The current version of the tool allows algebraic structures to be composed in up to 4 spatial dimensions. Formally, it implements the axioms of a \textit{quasistrict globular 4\-category}; the details of this theoretical basis are described by a subset of the authors in a corresponding theory paper~\cite{globular-theory}.

\globular is the first proof assistant of its kind, and it allows many proofs from higher category theory to be formalized, verified and visualized in a way that would not be practical in any other tool. The closest comparable tools are Quantomatic~\cite{quantomatic-cade}, which does diagrammatic rewriting for monoidal categories, Coq and Agda, which have been used to give formalizations of homotopy type theory~\cite{HoTTCoq}. The latter can indeed be used to perform logical and homotopy-theoretical proofs from a higher-categorical perspective; however, this approach diverges from ours in that it is based on the syntax of Martin-L\"of type theory rather than diagrams, and identity types naturally lead one to treat higher-dimensional \textit{invertible} structures (e.g. $\infty$-groupoids) as first-class citizens, rather than the more general structures we'll consider. Another comparable tool is \emph{Orchard}~\cite{orchard}, which allows the formalization of proofs in \textit{opetopic} (as opposed to \textit{globular}) higher categories; this tool can handle $\infty$-categories, and has many attractive properties, although the opetopic approach to higher categories is more restricted than the globular approach. The higher dimensional rewriting implemented by \globular draws inspiration from the polygraphic approach to rewriting \cite{Lafont2007,mimram-3d}, but extends it to allow for non-strict higher-categorical structures.

\globular was designed to make it as quick and easy as possible for users to go from zero to proving theorems and sharing proofs. Hence, it is entirely web-based, with all logic taking place client-side in the user's web browser. The most commonly-used procedures run in linear time with little overhead, so this is practical on modest hardware even for large diagrams. Proofs can be stored on the remote server for later reference, or downloaded for storage locally. Permanent hyperlinks to formalized proofs can be generated and embedded as links in research papers, allowing readers instant access to the formalization without the usual barriers-to-use of downloading, installing and maintaining an executable. The tool launched in December 2015, and has been well-received by the community, with 9055 sessions by 2052 unique users in under a year since deployment in December 2015\footnote{Usage statistics from Google Web Analytics retrieved on 28 November 2016.}.

In Section~\ref{sec:found}, we give a brief overview of the mathematical foundations of \globular, namely higher-dimensional category theory and rewriting. In Section~\ref{sec:using} we exhibit all of the core functionality of the tool via a simple example. In Section~\ref{sec:imp} we discuss the implementation, including the architecture and relevant data structures and procedures for rewriting. In Section~\ref{sec:technology} we describe some technical aspects of our implementation. In Section~\ref{sec:ex} we survey a variety of interesting proofs that have been formalized with \globular and made public, with direct links for viewing online.

This is an extended version of the conference paper~\cite{globular-fscd}. It has been updated to reflect the subsequent theoretical developments in~\cite{globular-theory}; in particular, regarding the fundamental algorithms, and the operation of the homotopy moves. A new example involving
Kan extensions and the codensity monad is also referenced.

\paragraph*{Acknowledgements} We would like to thank John Baez, Manuel B\"arenz, Bruce Bartlett, Eugenia Cheng, Chris Douglas, Eric Finster, Nick Gurski, Andr\'e Henriques, Samuel Mimram and Dominic Verdon for useful discussions.

\section{Mathematical foundations}\label{sec:found}

Higher category theory is the study of \textit{$n$-categories}. As well as objects and morphisms familiar from traditional category theory, which are called 0\-cells and 1\-cells, an $n$\-category also has morphisms between morphisms (2-cells), morphisms between those (3\-cells), and so on, up to level $n$. An $n$-category has a $n$ distinct composition operations, which allow cells to be combined to produce new cells.

\paragraph{Graphical calculus.}
A convenient notation for working with $n$-categories is the \textit{graphical calculus}, in which a $k$\-cell is represented as an $(n-k)$-dimensional geometrical structure\footnote{This is rigorously developed only for $n \leq 3$ \cite{barrett-graydiagrams,Joyal_1991}.}. Composition then  corresponds to `gluing' of these structures along the different axes of $n$-dimensional space. For example, in a 3\-category, we represent 3\-cells as points, 2\-cells as lines, 1\-cells as regions, and 0\-cells as `volumes'. Given 3\-cells $\alpha$ and $\beta$, we could form the following composite 3\-cells by composing along three different axes:
\begin{calign}
\label{eq:composites}
\begin{aligned}
\begin{tikzpicture}[thick, scale=0.8]
\draw [region] (-0.2,0) rectangle +(3,2);
\draw (0.8,0) to +(0,2);
\node [morphism] at (0.8,1) {$\beta$};
\draw [region] (0.0,-0.2) rectangle +(3,2);
\draw (2,-0.2) to +(0,2);
\node [morphism] at (2,0.8) {$\alpha$};
\end{tikzpicture}
\end{aligned}
&
\begin{aligned}
\begin{tikzpicture}[thick, scale=0.8]
\draw [region] (0,0) rectangle +(3,2);
\draw (1,0) to +(0,2);
\draw (2,0) to +(0,2);
\node [morphism] at (2,1) {$\beta$};
\node [morphism] at (1,1) {$\alpha$};
\end{tikzpicture}
\end{aligned}
&
\begin{aligned}
\begin{tikzpicture}[thick, scale=0.8]
\draw [region] (0,0) rectangle +(2,3);
\draw (1,0) to (1,2) to (1,3);
\node [morphism] at (1,2) {$\beta$};
\node [morphism] at (1,1) {$\alpha$};
\end{tikzpicture}
\end{aligned}
\end{calign}
In this way we can draw diagrams to represent arbitrary composites, in principle in any dimension; although for $n>3$, visualizing the resulting geometrical structure becomes nontrivial.

\paragraph{Rewriting.}
We take a \textit{rewriting} perspective on higher category theory. Suppose a $(k+1)$-cell $X$ has source and target $k$-cells $\alpha$ and $\alpha'$ respectively. Then we interpret $X$ as a way to rewrite $\alpha$ into $\alpha'$. Since composition in higher category theory is local, this also works for composite cells: for example, we can apply $X$ to any of the composites in \eqref{eq:composites} to obtain 
a new composite with $\alpha$ replaced by $\alpha'$.

The attractive feature of this perspective is that there is no fundamental difference between the notions of \textit{composition} and \textit{proof}. A proof that some diagram $D$ of $k$-cells can rewritten into some other diagram $D'$ amounts to building a composite $(k+1)$-cell with source $D$ and target $D'$, using just the `axiom' cells of a given theory. For instance, if we have a 3-cell called `\textsf{assoc}' which captures an associativity rule of 2-cells, we can prove a theorem about associativity as a composition of 3-cells:
\begin{center}
  \input{assoc3D.tikz}
\end{center}
That is, we can \emph{define} a composite $(k+1)$\-cell as a rewrite sequence on composite $k$-cells. This gives a recursive definition of composition, which terminates with a family of `basic' rewrite operations, which the user must specify. This is the essence of \globular's approach to higher category theory.



\paragraph{Strictness.}

In higher category theory, we have some freedom to decide what it means for two things to be `the same'. At one extreme are `fully weak' $n$-categories, where all of the axioms governing the composition of cells (such as  associativity and unit axioms) hold only up to higher-dimensional cells. For example, for $1$-cells $f, g, h$, rather than requiring associativity
$$f \circ (g \circ h) = (f \circ g) \circ h$$
we merely assert the existence of a (weakly) invertible family of `associator' $2$-cells
$$(f \circ g) \circ h \to f \circ (g \circ h).$$
These in turn must satisfy various coherence properties, which we again interpret only up to higher-dimensional cells (which \textit{themselves} must satisfy coherence properties, and so on). While these structures arise naturally in many contexts, the substantial bureaucracy that arises from this structure makes it hard to work with weak $n$-categories as purely syntactic objects. 

At the other extreme are the \textit{strict $n$-categories} which require all the axioms involving composition of cells to hold as on-the-nose equalities. These are quite easy to define~\cite{leinster-survey}, and admit an evident notion of finite presentation, called a \textit{polygraph} or \textit{computad}, and have a reasonably well-behaved \textit{higher-dimensional rewrite theory} \cite{guiraud-termination}. However, for $n>2$, it is \textit{not} the case that every weak $n$-category is equivalent to a strict one.

\paragraph{Homotopies.}
To see where this richness of weak categories comes from, we consider the interchange law, which in a 2\-category acts as follows as a rewrite on composite 2\-cells:
\begin{center}
$
({\color{blue} \bm{f}} \circ_1 1_B) \circ_2 (1_{A'} \circ_1 {\color{red} \bm{g}}) \ \ \rightarrow\ \ (1_{A} \circ_1 {\color{red} \bm{g}}) \circ_2 ({\color{blue} \bm{f}} \circ_1 1_{B'})
$

$
\begin{aligned}
\begin{tikzpicture}[thick, yscale=0.8]
\draw (0,0.2) to (0,2.75);
\draw (1,0.2) to (1,2.75);
\node [blue dot] at (0,1) {};
\node [red dot] at (1,2) {};
\end{tikzpicture}
\end{aligned}
\quad\xrightarrow{\ \ \ I\ \ \ }\quad
\begin{aligned}
\begin{tikzpicture}[thick, yscale=0.8]
\draw (0,0.2) to (0,2.75);
\draw (1,0.2) to (1,2.75);
\node [blue dot] at (0,2) {};
\node [red dot] at (1,1) {};
\end{tikzpicture}
\end{aligned}
$
\end{center}
When we stop at two dimensions, there is no problem treating this `node-sliding' rule simply as an equation between diagrams. 
But seen as a 3\-cell in a 3\-category, the source and target of $I$ become the bottom and top slice of a 3D picture, the nodes become wires, and the `sliding' becomes a braiding:
\begin{center}
$
\begin{aligned}
\begin{tikzpicture}[thick]
\draw [region] (0,0) rectangle (2,2);
\draw [red] (0.6,0) to [out=up, in=down] (1.6,1.8) to +(0,0.2);
\begin{scope}[yshift=-0.2cm, xshift = 0.2cm]
\draw [region] (0,0) rectangle +(2,2);
\draw [blue] (1.4,0) to [out=up, in=down] (0.4,2);
\end{scope}
\end{tikzpicture}
\end{aligned}
$
\end{center}
By the invertibility and naturalness properties, these braidings then behave exactly how you would expect genuine topological braids to behave. For instance, the following higher rewrites exist:
\begin{center}
\tikzset{every picture/.style={scale=0.8}}
$
\begin{aligned}
\begin{tikzpicture}[thick, yscale=0.7]
\draw [region] (0,0) rectangle (2,4);
\draw [red] (0.6,0) to [out=up, in=down] (1.6,1.9) to [out=up, in=down] (0.6,3.8) to +(0,0.2);
\draw [region] (0.2,-0.2) rectangle +(2,4);
\draw [blue] (1.6,-0.2) to +(0,0.2) to [out=up, in=down] (0.6,1.9) to [out=up, in=down] (1.6,3.8);
\end{tikzpicture}
\end{aligned}
\quad\begin{array}{@{}c@{}}\rightarrow\\\leftarrow\end{array}\quad
\begin{aligned}
\begin{tikzpicture}[thick, yscale=0.7]
\draw [region] (0,0) rectangle (2,4);
\draw [red] (0.6,0) to (0.6,3.8) to +(0,0.2);
\draw [region] (0.2,-0.2) rectangle +(2,4);
\draw [blue] (1.6,-0.2) to (1.6,3.8);
\end{tikzpicture}
\end{aligned}
\qquad\quad
\ignore{\begin{aligned}
\begin{tikzpicture}[thick, yscale=0.7]
\draw [region] (0,0) rectangle (2,4);
\draw [red] (1.6,0) to (1.6,4);
\draw [region] (0.2,-0.2) rectangle +(2,4);
\draw [blue] (0.6,-0.2) to (0.6,3.8);
\end{tikzpicture}
\end{aligned}
\quad\begin{array}{@{}c@{}}\rightarrow\\\leftarrow\end{array}\quad
\begin{aligned}
\begin{tikzpicture}[thick, yscale=0.7]
\draw [region] (0,0) rectangle (2,4);
\draw [red] (1.6,0) to [out=up, in=down] (0.6,1.9) to [out=up, in=down] (1.6,3.8) to +(0,0.2);
\draw [region] (0.2,-0.2) rectangle +(2,4);
\draw [blue] (0.6,-0.2) to +(0,0.2) to [out=up, in=down] (1.6,1.9) to [out=up, in=down] (0.6,3.8);
\end{tikzpicture}
\end{aligned}}
\def\off{0.4}
\begin{aligned}
\begin{tikzpicture}[thick, yscale=0.55]
\draw [region, fill opacity=0.7] (-1.7,0) rectangle +(3,6+2*\off);
\draw [red] (-1,0) to [out=up, in=down] (0,2) to [out=up, in=down] (1,4) to (1,6+2*\off);
\draw [region, fill opacity=0.7] (-1.5,-\off) rectangle +(3,6+2*\off);
\draw [blue] (0,-\off) to (0,0) to [out=up, in=down] (-1,2) to (-1,4) to [out=up, in=down] (0,6) to (0,6+\off);
\draw [region, fill opacity=0.7] (-1.3,-2*\off) rectangle +(3,6+2*\off);
\draw [green!60!black] (1,-2*\off) to (1,2) to [out=up, in=down] (0,4) to [out=up, in=down] (-1,6) to (-1,6);
\end{tikzpicture}
\end{aligned}
\quad\begin{array}{@{}c@{}}\rightarrow\\\leftarrow\end{array}\quad
\begin{aligned}
\begin{tikzpicture}[thick, yscale=0.55]
\draw [region, fill opacity=0.7] (-1.7,0) rectangle +(3,6+2*\off);
\draw [red] (-1,0) to [out=up, in=down] (-1,2) to [out=up, in=down] (0,4) to [out=up, in=down] (1,6) to (1,6+2*\off);
\draw [region, fill opacity=0.7] (-1.5,-\off) rectangle +(3,6+2*\off);
\draw [blue] (0,-\off) to (0,0) to [out=up, in=down] (1,2) to (1,4) to [out=up, in=down] (0,6) to (0,6+\off);
\draw [region, fill opacity=0.7] (-1.3,-2*\off) rectangle +(3,6+2*\off);
\draw [green!60!black] (1,-2*\off) to (1,0) to [out=up, in=down] (0,2) to [out=up, in=down] (-1,4) to (-1,6);
\end{tikzpicture}
\end{aligned}
$
\end{center}
In general, overcrossings and undercrossings are distinct, so it is possible for wires to become tangled. Requiring interchangers to be identities, as in the theory of strict 3\-categories, trivializes this part of the theory, and means that it is no longer fully general, in the precise sense that not every 3\-category is equivalent to a strict 3\-category.

It follows that the strict $n$-categorical setting in which the polygraph community work is not sufficiently general to reason about arbitrary $n$-categories. The solution is to work instead with \textit{semistrict} $n$-categories, which allows a small amount of weak structure, sufficient to ensure that every weak n\-category is equivalent to semistrict $n$\-category. For $n=3$, \textit{Gray categories} have this property; they are defined as 3\-categories in which all weak structure is the identity, except for interchangers\footnote{A definition of semistrict $n$-category for $n>3$ has not yet been generally accepted.}. For $n=4$, \globular implements a new definition of \textit{quasistrict 4\-categories}~\cite{globular-theory}, which can be considered 4\-dimensional generalizations of Gray categories.


\section{Using \globular}
\label{sec:using}

Constructing a theory and proving theorems in \globular is an inductive process, whereby lower-dimensional objects are used to construct higher-dimensional objects. This is done by building up a \textit{signature}, i.e. a collection of generators, in parallel with increasingly higher-dimensional diagrams. From an empty signature, the only thing to do is add new $0$-cells:
\begin{center}
  \includegraphics[height=2cm]{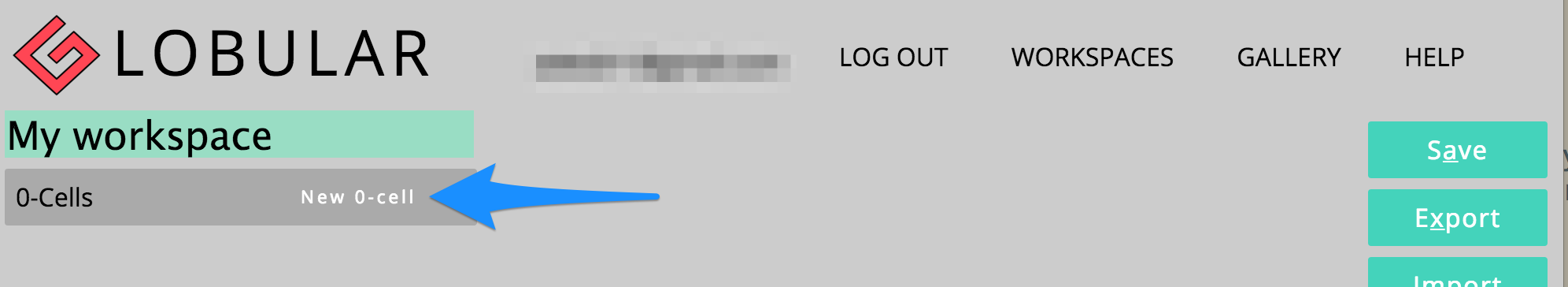}
\end{center}
Once we have some $0$-cells, these can be made the sources and targets of new $1$-cells:
\begin{center}
$
  \begin{aligned}
  \includegraphics[height=3cm]{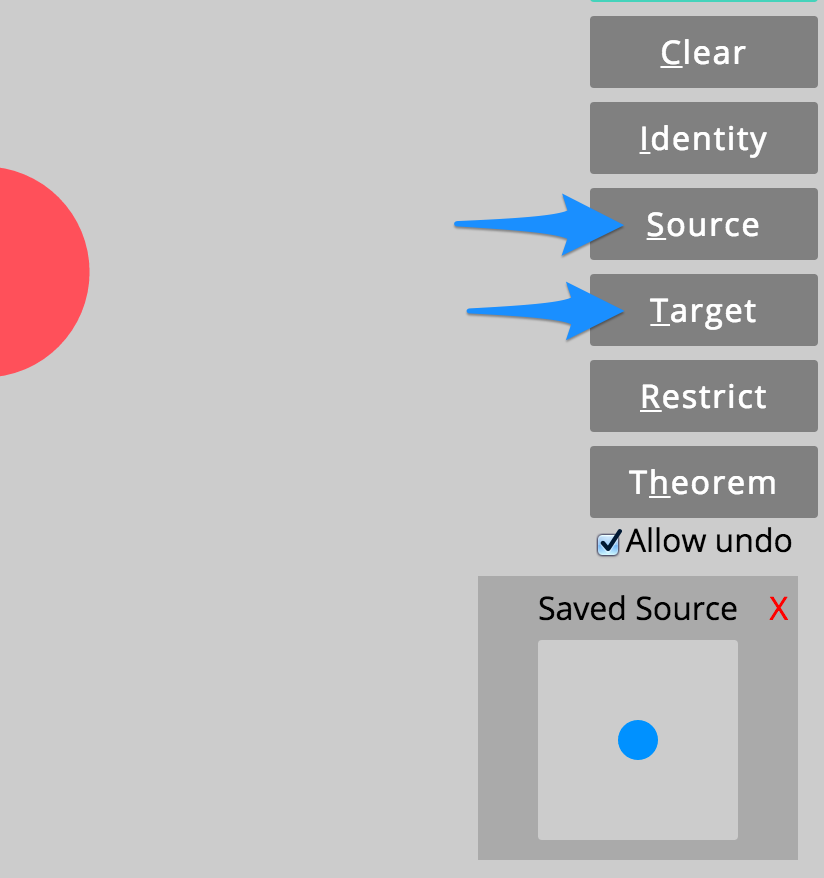}
  \end{aligned}
  \ \ \Rightarrow\ \ 
  \begin{aligned}
  \includegraphics[height=3cm]{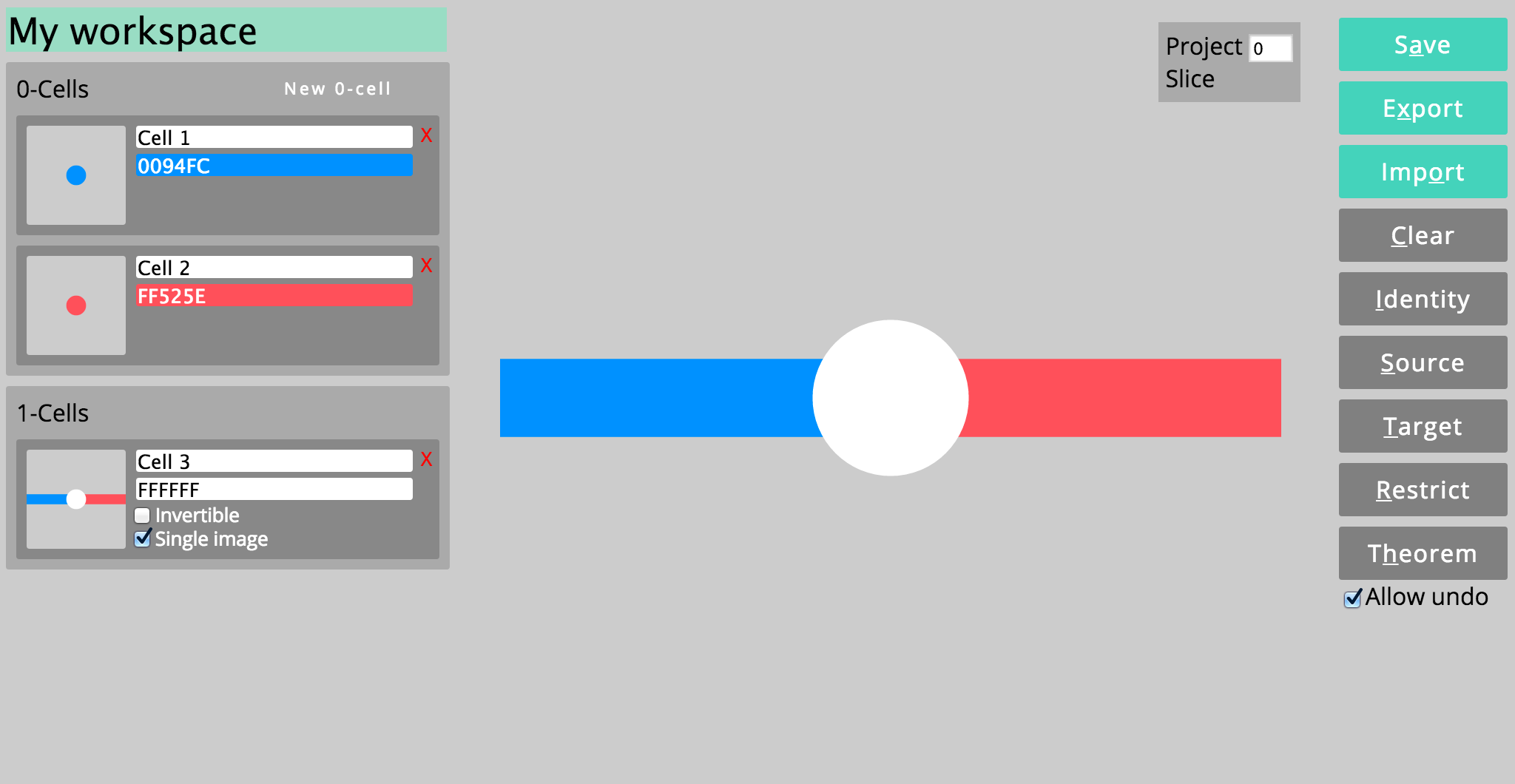}
  \end{aligned}
$
\end{center}
At this point things start to get interesting, since $1$-cells can be \textit{attached} to each other to form non-trivial diagrams. These diagrams can then form the sources and targets of new $2$-cells:
\begin{center}
$
\begin{aligned}
  \includegraphics[height=2cm]{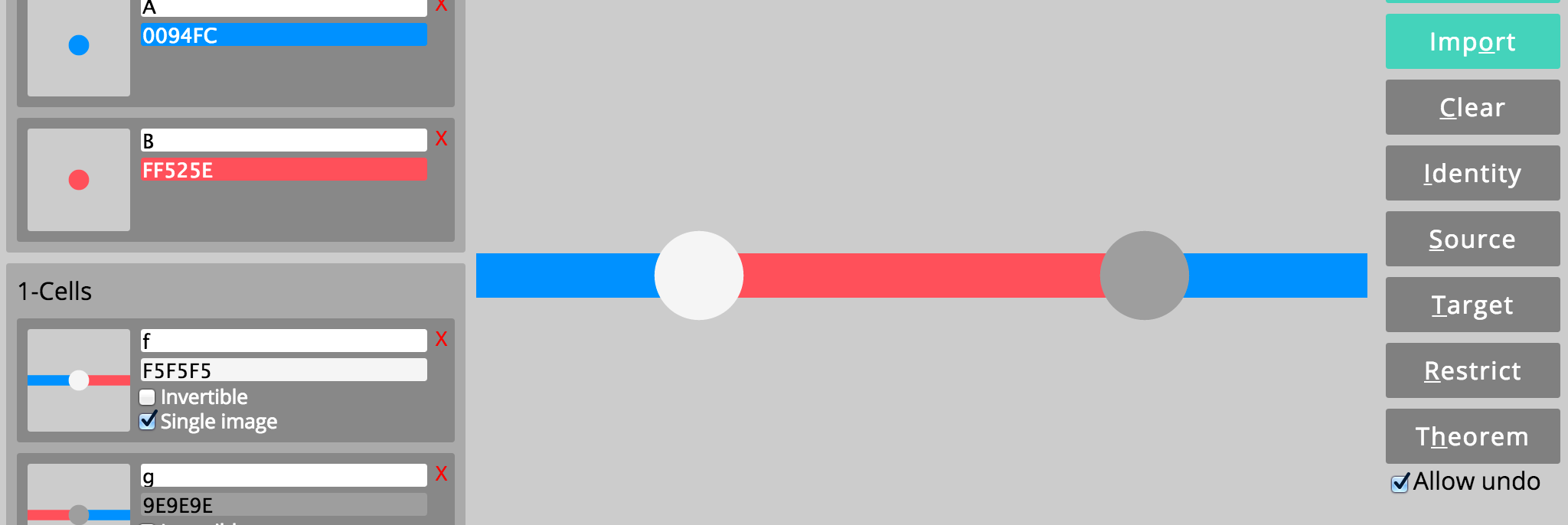}
\end{aligned}
\ \ \Rightarrow\ \ 
\begin{aligned}
\includegraphics[width = 60pt, height = 30pt]{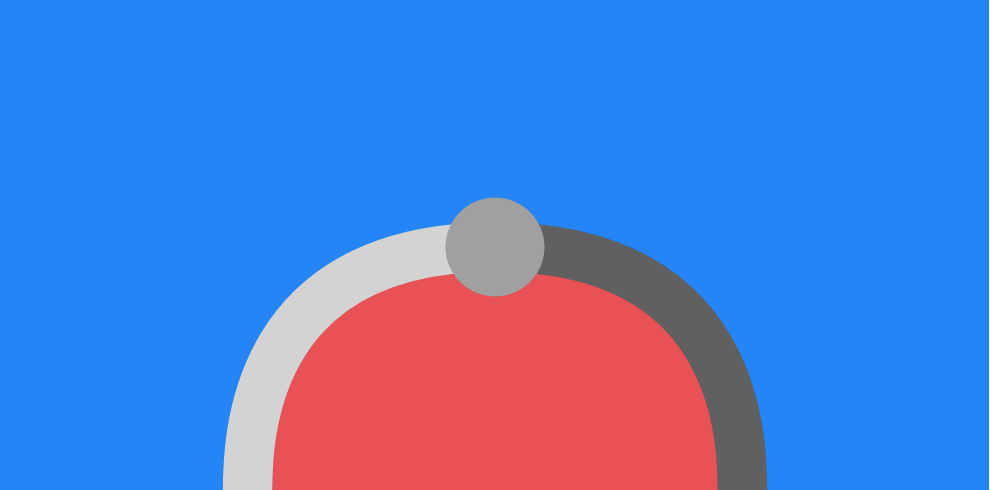}
\end{aligned}
\ \ ,\ \ 
\begin{aligned}
\includegraphics[width = 60pt, height = 30pt]{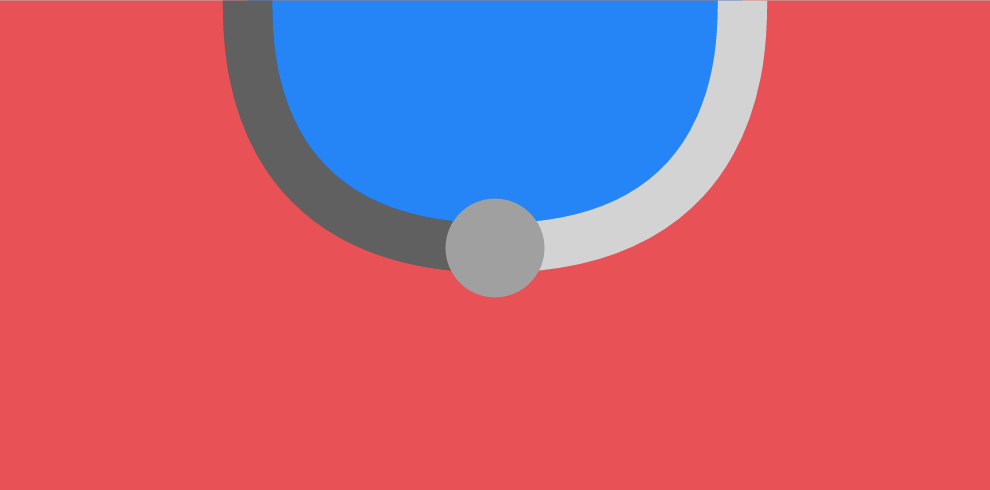}
\end{aligned}
$
\end{center}
In turn, these $2$-cells can be composed to form larger diagrams, which and form the sources and targets of new $3$-cells. We can either interpret these new $3$-cells as new generators, or as \textit{equations} between 2d diagrams. For example, we can make our `cap' and `cup' $2$-cells invertible by adding the following $3$-cells to our theory:
\begin{center}
$
\begin{aligned}
\includegraphics[width = 60pt, height = 60pt]{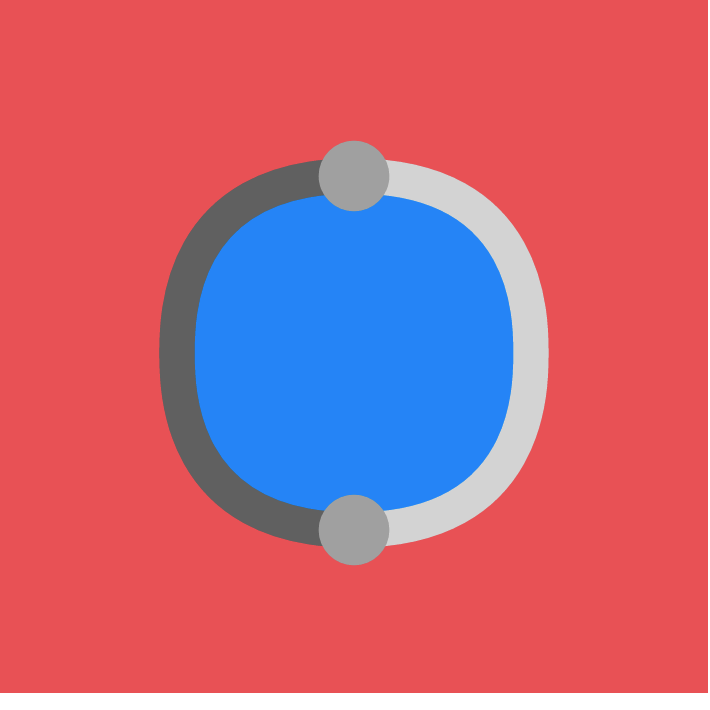}
\end{aligned}
\quad\begin{array}{@{}c@{}}\pi_1\\\rightarrow\\\leftarrow\\\pi_2\end{array}\quad
\begin{aligned}
\includegraphics[width = 60pt, height = 60pt]{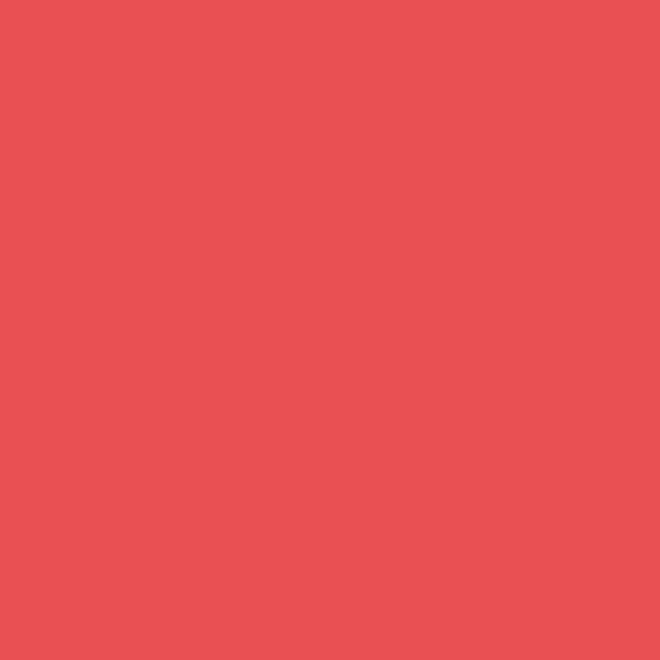}
\end{aligned}
\qquad\qquad
\begin{aligned}
\includegraphics[width = 60pt, height = 60pt]{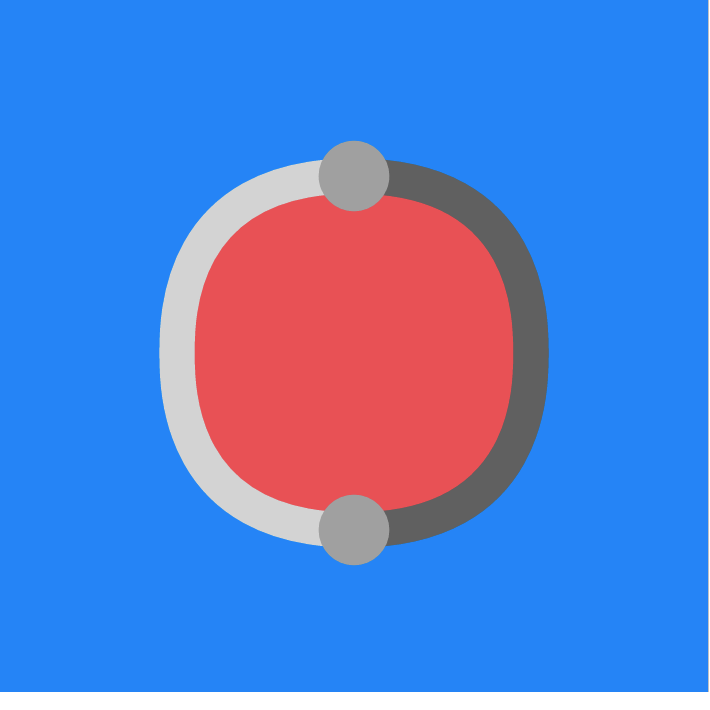}
\end{aligned}
\quad\begin{array}{@{}c@{}}\pi_3\\\rightarrow\\\leftarrow\\\pi_4\end{array}\quad
\begin{aligned}
\includegraphics[width = 60pt, height = 60pt]{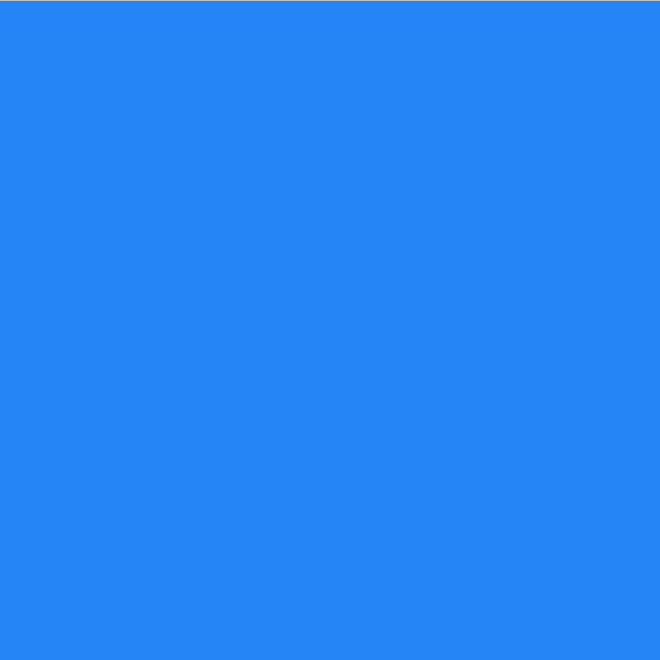}
\end{aligned}
$
\end{center}

\begin{center}
$
\begin{aligned}
\includegraphics[width = 60pt, height = 60pt]{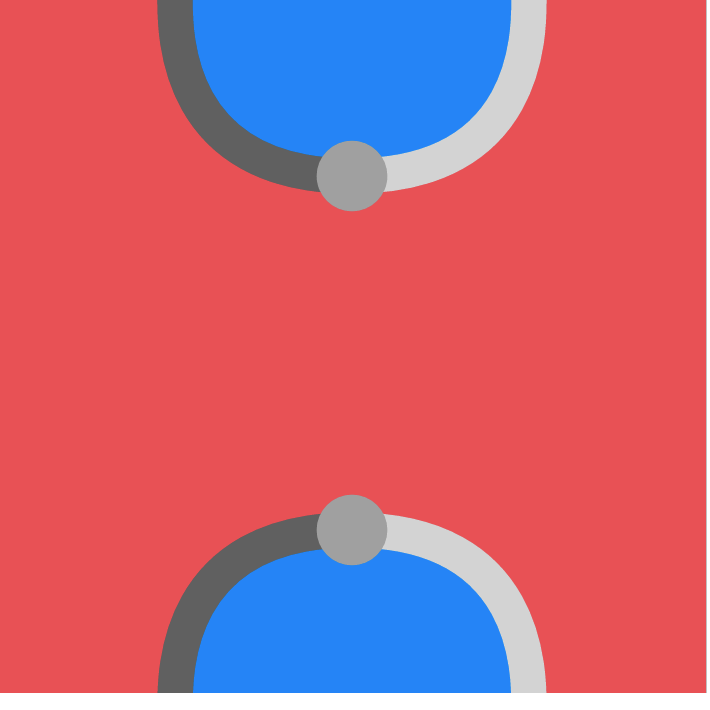}
\end{aligned}
\quad\begin{array}{@{}c@{}}\pi_5\\\rightarrow\\\leftarrow\\\pi_6\end{array}\quad
\begin{aligned}
\includegraphics[width = 60pt, height = 60pt]{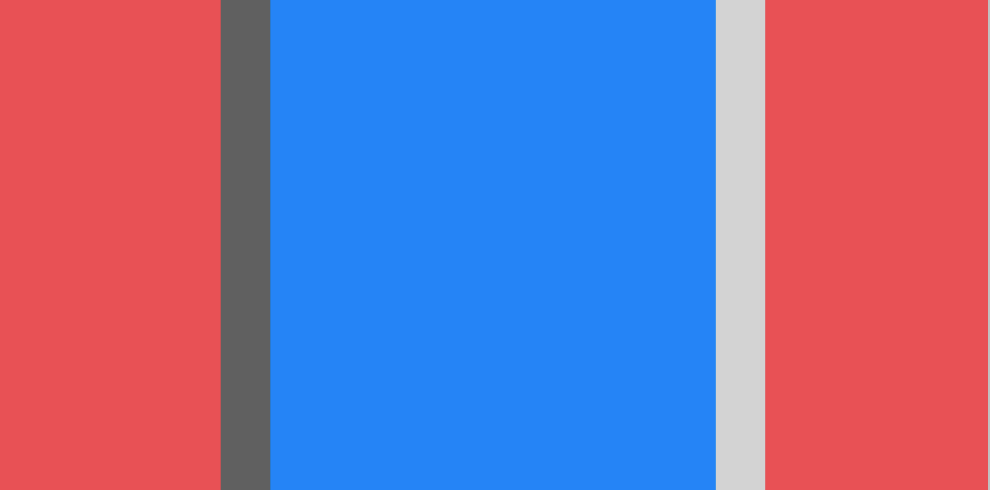}
\end{aligned}
\qquad\qquad
\begin{aligned}
\includegraphics[width = 60pt, height = 60pt]{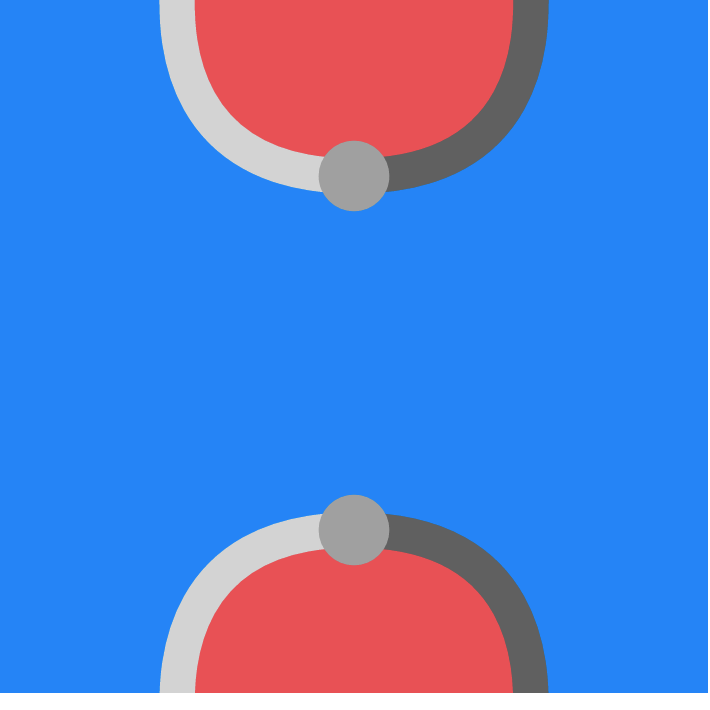}
\end{aligned}
\quad\begin{array}{@{}c@{}}\pi_7\\\rightarrow\\\leftarrow\\\pi_8\end{array}\quad
\begin{aligned}
\includegraphics[width = 60pt, height = 60pt]{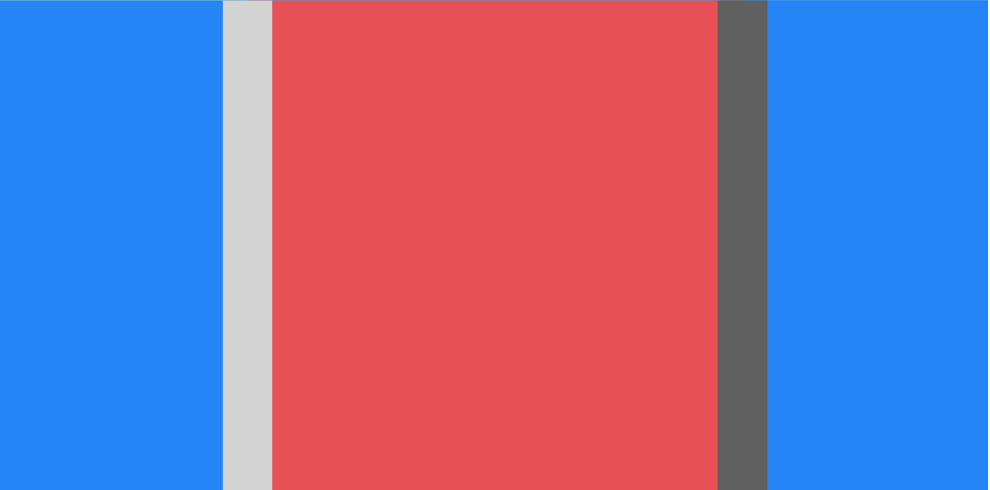}
\end{aligned}
$
\end{center}
These invertible `cup' and `cap' 2-cells yield a familiar categorical structure.
\begin{definition}
In a 2-category, an \textit{equivalence} is a pair of objects $A$ and $B$, a pair of $1$-cells $A \xrightarrow{F} B$ and $B \xrightarrow{G} A$ and invertible $2$-cells $F\circ G \xrightarrow{\alpha} \text{id}_A$ and $\text{id}_A \xrightarrow{\beta} G\circ F$, denoted as follows:
\begin{center}
$
\alpha\equiv
\begin{aligned}
\includegraphics[width = 60pt, height = 30pt]{Equivalence-19.png}
\end{aligned}
\qquad
\beta\equiv
\begin{aligned}
\includegraphics[width = 60pt, height = 30pt]{Equivalence-20.png}
\end{aligned}
$
\end{center}
\end{definition}

A special case is where the 2-category is \textbf{Cat}, in which case this yields the usual notion of equivalence of categories. Then the following is a well-known fact about equivalences in a 2-category~\cite{baezlauda, Rivano}:

\begin{theorem}
In a 2-category, every equivalence gives rise to a dual equivalence.
\end{theorem}
An equivalence is called a \textit{dual equivalence} if it additionally satisfies the \textit{snake equations}, which take the following geometrical form:
\begin{align}
\begin{aligned}\includegraphics{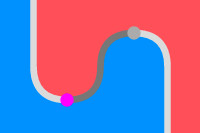}
\end{aligned}
&=
\begin{aligned}
\includegraphics{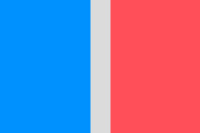}
\end{aligned}
&
\begin{aligned}\includegraphics{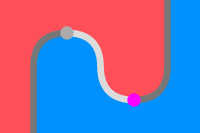}
\end{aligned}
&=
\begin{aligned}
\includegraphics{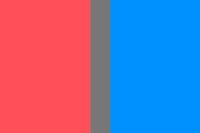}
\end{aligned}
\end{align}

\noindent We can prove these theorems by replacing the `cup' with a `sock', defined in terms of the old cup and cap:
\begin{center}
  \includegraphics[height=2cm]{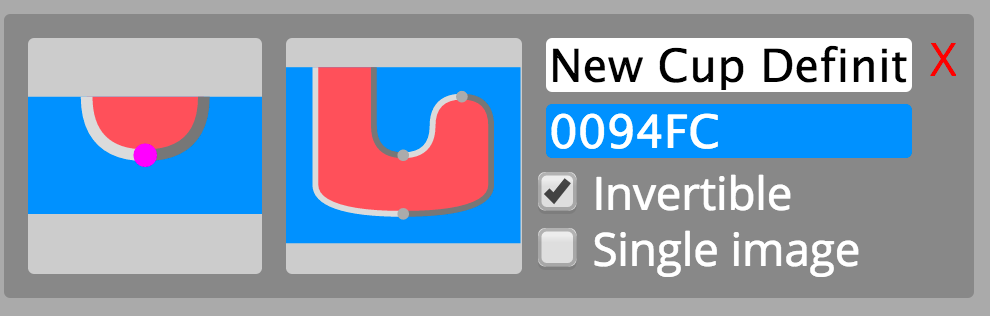}
\end{center}
We can show that our new `cup' satisfies the snake equation, with the original `cap'. To prove the first snake equation, we perform the following non-trivial sequence of rewrites in \globular, where $\sim$ indicates a \textit{homotopy move} discussed later in Section~\ref{sec:homotopy}:
\begin{flalign*}
&
\begin{aligned}
\includegraphics[width = 45pt, height = 40pt]{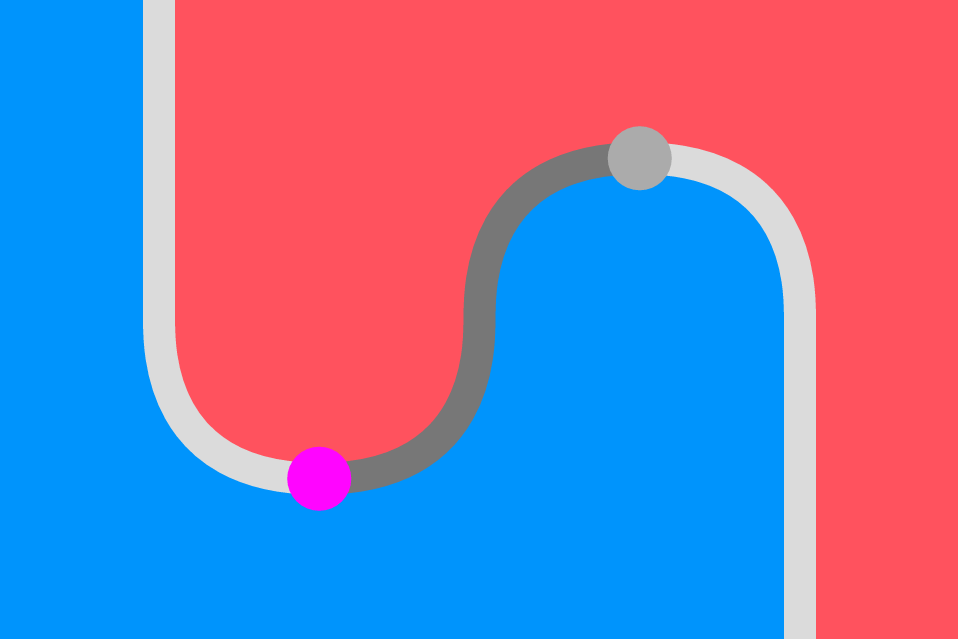}
\end{aligned}
\quad \stackrel {\text{def}} \to \quad
\begin{aligned}
\includegraphics[width = 45pt, height = 40pt]{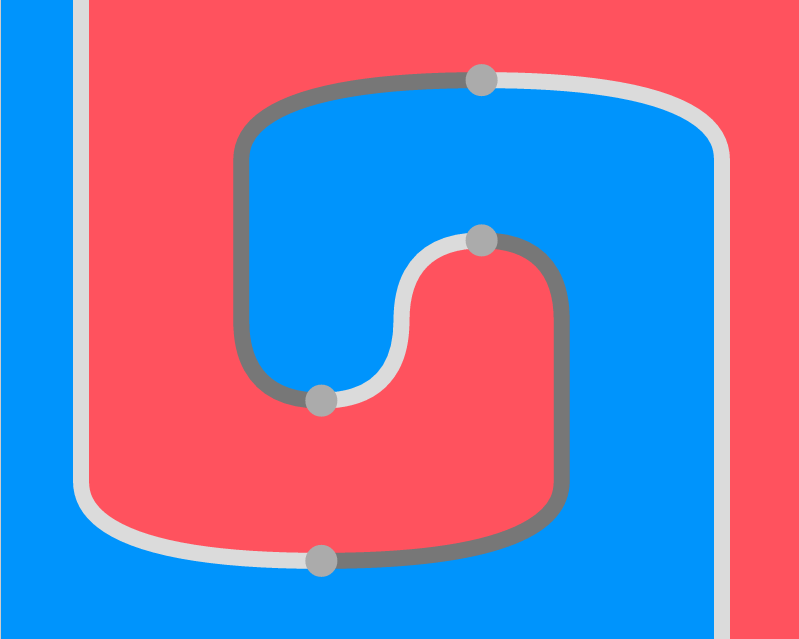}
\end{aligned}
\quad\stackrel {\pi_2} \to\quad
\begin{aligned}
\includegraphics[width = 45pt, height = 40pt]{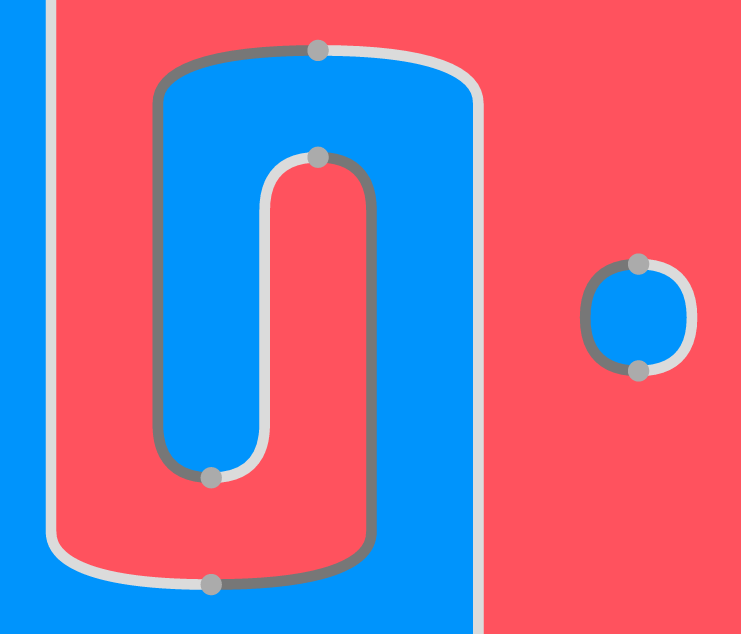}
\end{aligned}
\quad\stackrel {\pi_8} \to\quad
\begin{aligned}
\includegraphics[width = 45pt, height = 40pt]{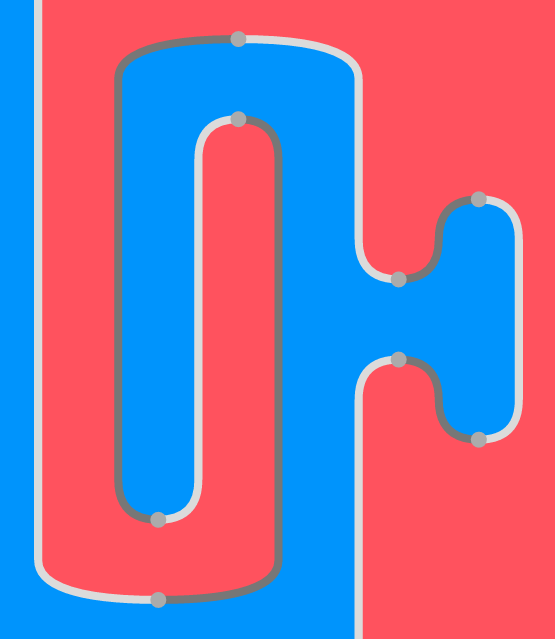}
\end{aligned}
\quad\stackrel {\sim} \to\quad
\begin{aligned}
\includegraphics[width = 45pt, height = 40pt]{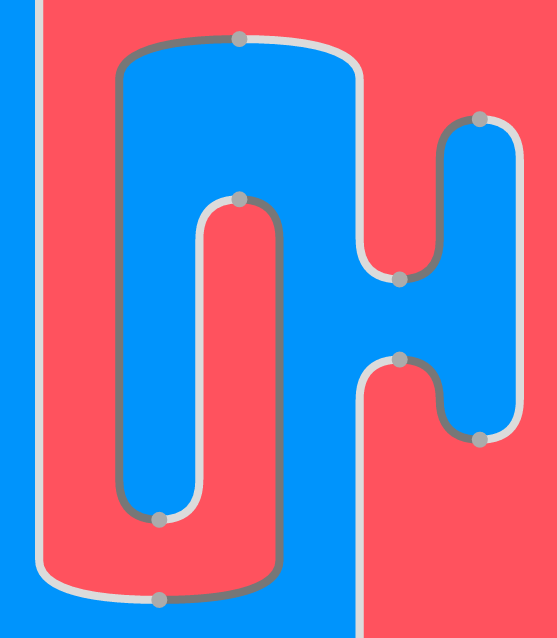}
\end{aligned}
\quad\stackrel {\sim} \to\quad
& \\ &
\begin{aligned}
\includegraphics[width = 45pt, height = 40pt]{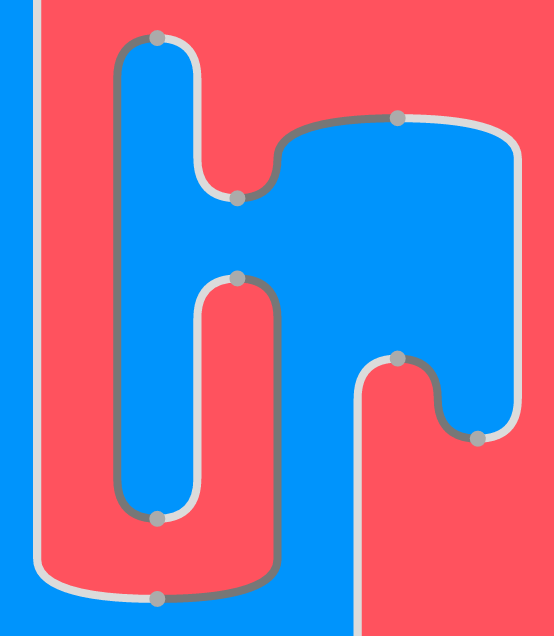}
\end{aligned}
\quad\stackrel {\sim} \to\quad
\begin{aligned}
\includegraphics[width = 45pt, height = 40pt]{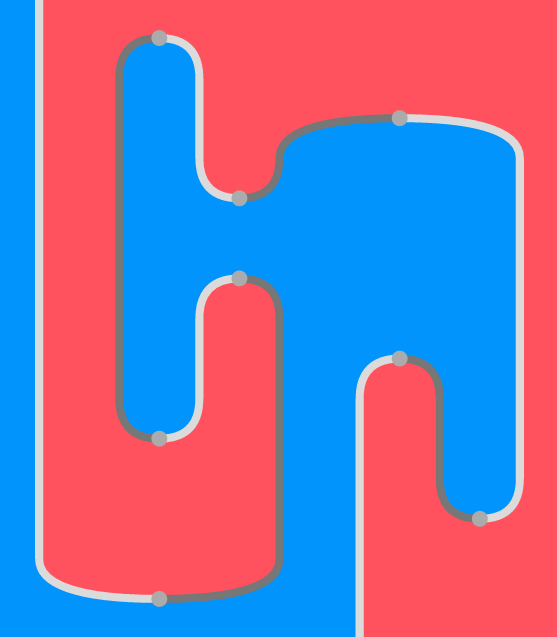}
\end{aligned}
\quad\stackrel {\sim} \to\quad
\begin{aligned}
\includegraphics[width = 45pt, height = 40pt]{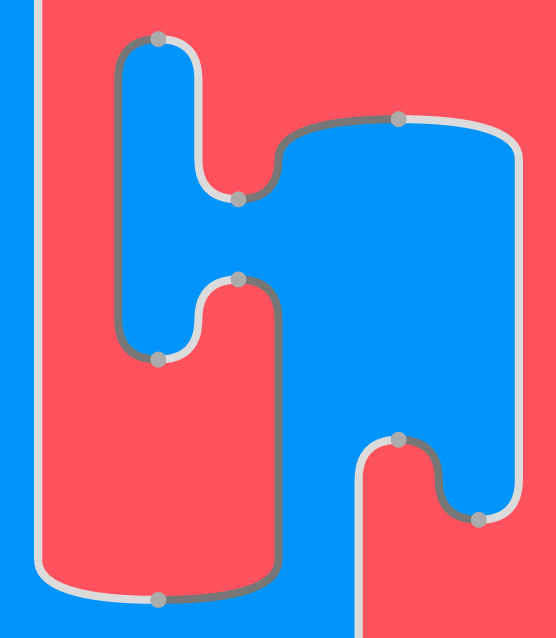}
\end{aligned}
\quad\stackrel {\sim} \to\quad
\begin{aligned}
\includegraphics[width = 45pt, height = 40pt]{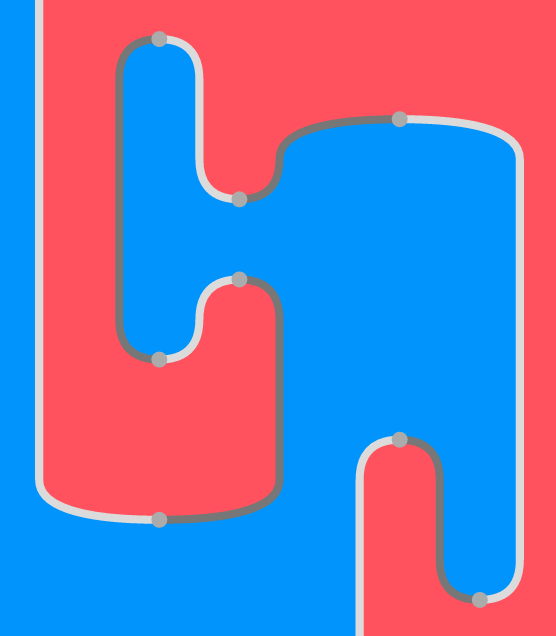}
\end{aligned}
\quad\stackrel {\sim} \to\quad
\begin{aligned}
\includegraphics[width = 45pt, height = 40pt]{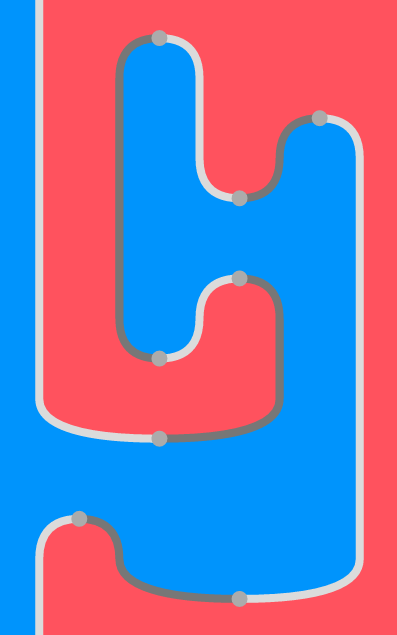}
\end{aligned}
\quad\stackrel {\pi_7} \to\quad
& \\ &
\begin{aligned}
\includegraphics[width = 45pt, height = 40pt]{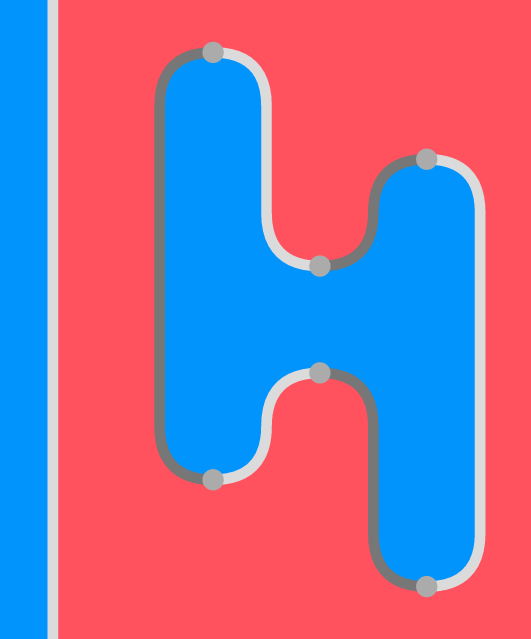}
\end{aligned}
\quad\stackrel {\pi_7} \to\quad
\begin{aligned}
\includegraphics[width = 45pt, height = 40pt]{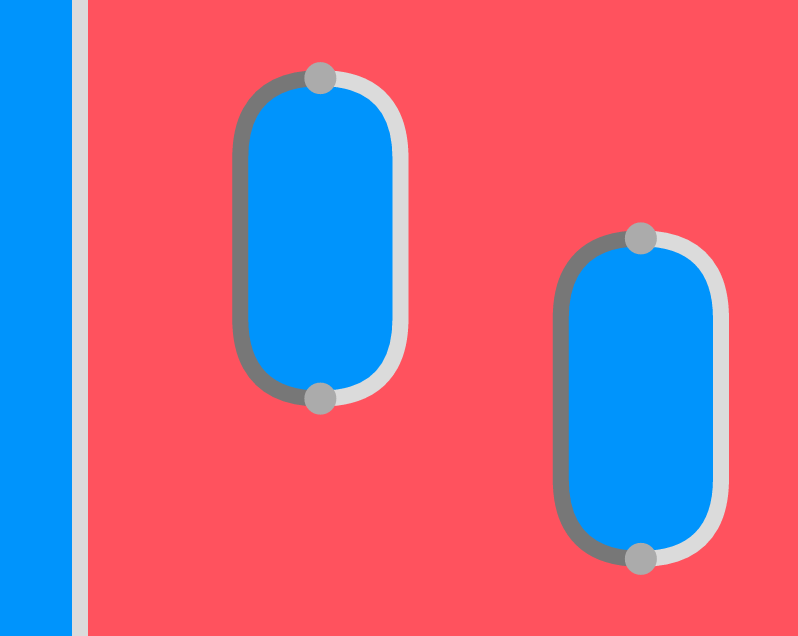}
\end{aligned}
\quad\stackrel {\sim} \to\quad
\begin{aligned}
\includegraphics[width = 45pt, height = 40pt]{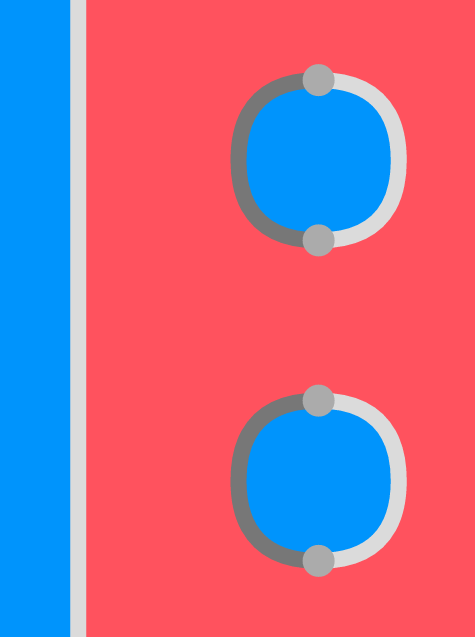}
\end{aligned}
\quad\stackrel {\pi_1} \to\quad
\begin{aligned}
\includegraphics[width = 45pt, height = 40pt]{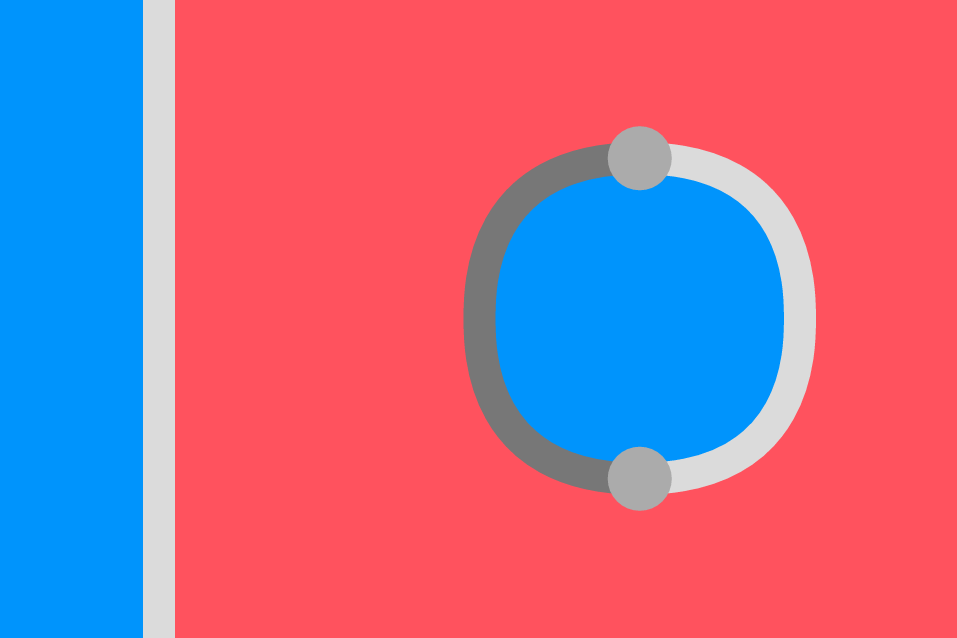}
\end{aligned}
\quad\stackrel {\pi_1} \to\quad
\begin{aligned}
\includegraphics[width = 45pt, height = 40pt]{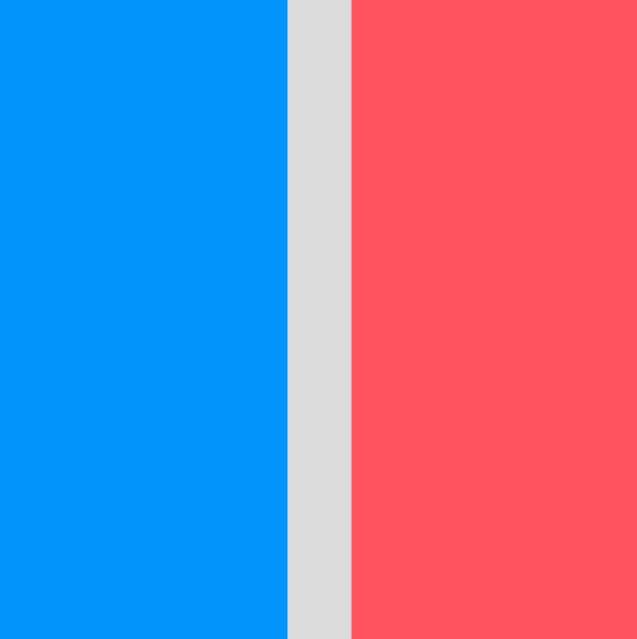}
\end{aligned}
&
\end{flalign*}
This proof is itself a 3-cell, and is represented by a single 3\-dimensional object. In \globular, we can either browse through it slice-by-slice, or we can see the overall structure of the proof as a single diagram, by choosing `Project=1' in the interface:
\begin{center}
\includegraphics[width = 200pt]{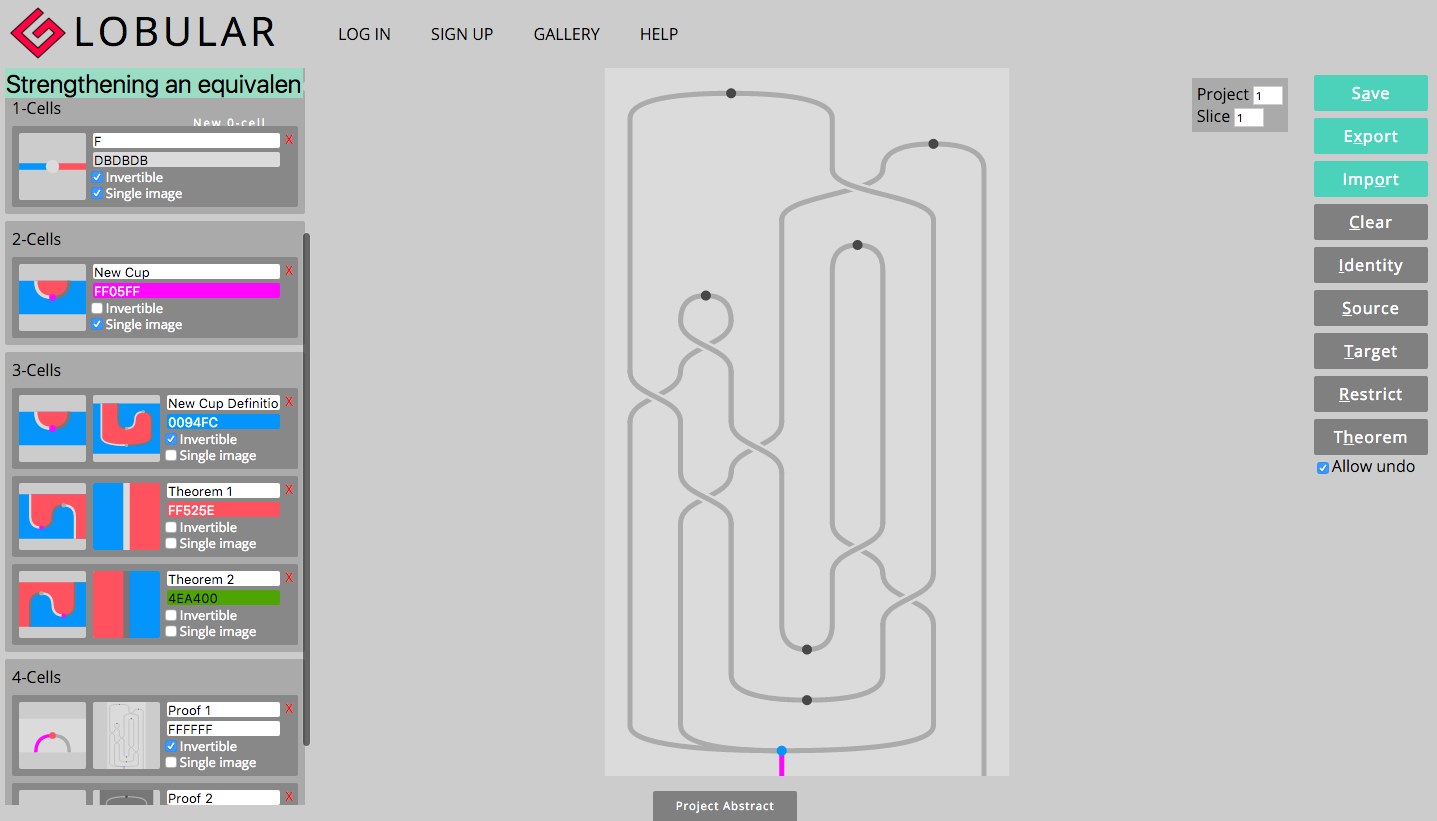}
\end{center}
This projects out one dimension so we call look at this entire 3-cell `side-on'. The nodes represent applications of rewrite rules, and the wires represent $2$-cells. From this view, we can refactor the proof by eliminating redundant steps (e.g. a rewrite immediately followed by its inverse) or by re-ordering rewrites that are applied to independent parts of the diagram.

Once a proof has been constructed, it can be saved privately to the server, or made public by \textit{publishing} it. This assigns the workspace a permanent unique link, which can be shared with others or linked from a research paper. For example, the proof in this section is based on the formalization available here: \hrefLMCS{http://globular.science/1512.007}{globular.science/1512.007}.

\section{Data structures and algorithms}\label{sec:imp}

\subsection{Data structures}

The fundamental structures that \globular makes use of are \emph{signatures}, which are a lists of basic generating cells that the user has specified, and \emph{diagrams}, which are particular composites of generators from a given signature. These can be defined compactly in a mutually-recursive fashion. For clarity, we write these both as type families in dependent type-style notion. Let `$g : \Set$' declare a finite set $g$ (which we then treat as a type), let $\List(\alpha)$ be the type of lists, and $\dpair{\alpha_1, \alpha_2, \ldots}$ the type of tuples where types in $\alpha_j$ are allowed to depend on $\alpha_i$ for $i < j$. Let $\Sig(0)$ and $\Diag(0, *)$ both be the unit type $\{*\}$. Then, for $n > 0$:
\begin{center}
  $
  \begin{array}{l}
  \Sig(n : \mathbb N) := \\
  \qquad\dpair{\begin{array}{l}
    g : \Set, \\
    \sigma : \Sig(n - 1), \\
    s, t : g \to \Diag(n-1, \sigma)
  \end{array}
  }\end{array}
  \qquad
  \begin{array}{l}
\Diag(n : \mathbb N, \sigma : \Sig(n)) := \\
  \qquad\dpair{\begin{array}{l}
    s : \Diag(n-1,\sigma), \\
    \delta : \List( \ \dpair{ a: \sigma.g,\  c: \List(\N) } \  )
  \end{array}
  }
\end{array}
  $
\end{center}

\noindent An $n$\-signature $\Sigma: \Sig(n)$ therefore consists of an $(n-1)$\-signature $\Sigma.\sigma$, and a set of generators $\Sigma.g$, such that each $x : \Sigma.g$ has a source and target $(n-1)$-diagrams $\Sigma.s(x)$ and $\Sigma.t(x)$ respectively, which each contain cells from the $(n-1)$\-signature $\Sigma.\sigma$.

Given a signature $\sigma: \Sig(n)$, then for $n>0$, a diagram $\Delta : \Diag(n,\sigma)$ consists of a source $(n-1)$-diagram $\Delta.s$, and a list of $n$-cells $\Delta_\delta$ that act sequentially on that source. The $k$th $n$-cell is given by a pair $\Delta.\delta[k]$, whose first element $\Delta.\delta[k].a$ is a generating cell drawn from the signature $\sigma$, and whose second element $\Delta.\delta[k].c$ is a list of $n{-}1$ numbers which specify the coordinates at which the chosen generating rewrite acts. For example, a $2$-diagram consists of a list of $2$-cells which are stacked vertically, and this coordinate consists of a single number giving the horizontal position of each $2$-cell; that is, the number of wires appearing to its left. In general, the coordinates give the `height' of the rewrite in every dimension, starting from the top-dimensional one. We leave the target $(n-1)$-cell implicit, as it can be recovered from the other data (e.g. via the \textbf{Slice} procedure below).

We illustrate this informally with the following example. Let $\Sigma : \Sig(2)$ be a signature containing the following generators:
\begin{align*}
\Sigma.g &= \{v_1, v_2, v_3, v_4, v_5\}
\\
\Sigma.\sigma.g &= \{e_1, e_2\}
\\
\Sigma.\sigma.\sigma.g &= \{r_1, r_2\}
\end{align*}
Then there is source and target data for these generators such that the following diagrams can all be constructed:
\begin{align*}
\begin{aligned}
\begin{array}{l@{\hspace{3pt}}l}
D_1.\delta &= \big[ (v_1, [0]), (v_3, [1]), (v_2, [1]) \big]
\\[3pt]
D_1.s.\delta &= \big[(e_1, []) \big]
\end{array}
\end{aligned}
\qquad
\begin{aligned}
\begin{tikzpicture}[thick, scale=0.7]
\draw (0,0) to (0,1) to [out=right, in=down] (0.5,1.5) to (0.5,2);
\draw (0,1) to [out=left, in=down] (-0.5,1.5) to (-0.5,4);
\draw (0.5,3) to (0.5,4);
\node [blob] at (0.5,3) {};
\node [blob] at (0,1) {};
\node [blob] at (0.5,2) {};
\node [label] at (-1.4,2) {$r_1$};
\node [label] at (1.7,2) {$r_2$};
\node [label, left] at (-0.5,3) {$e_1$};
\node [label, left] at (0,0.56) {$e_1$};
\node [label, right] at (0.5,1.5) {$e_2$};
\node [label, right] at (0.5,3.4) {$e_2$};
\node [label, below right] at (0,1.15) {$v_1$};
\node [label, right] at (0.5,2.85) {$v_2$};
\node [label, right] at (0.5,2.15) {$v_3$};
\end{tikzpicture}
\end{aligned}
\end{align*}

\begin{align*}
\begin{aligned}
\begin{array}{l@{\hspace{3pt}}l}
D_2.\delta  &= \big[ (v_1, [0]), (v_2, [2]), (v_3, [1]) \big]
\\[3pt]
D_2.s.\delta &= \big[(e_1, []) \big]
\end{array}
\end{aligned}
\qquad
\begin{aligned}
\begin{tikzpicture}[thick, scale=0.7]
\draw (0,0) to (0,1) to [out=right, in=down] (0.5,1.5) to (0.5,3);
\draw (0,1) to [out=left, in=down] (-0.5,1.5) to (-0.5,4);
\draw (1.5,2) to (1.5,4);
\node [blob] at (1.5,2) {};
\node [blob] at (0,1) {};
\node [blob] at (0.5,3) {};
\node [label] at (-1.2,2) {$r_1$};
\node [label] at (2.3,2) {$r_2$};
\node [label, left] at (-0.5,3) {$e_1$};
\node [label, left] at (0,0.56) {$e_1$};
\node [label, left] at (0.5,2) {$e_2$};
\node [label, right] at (1.5,3) {$e_2$};
\node [label, below right] at (0,1.15) {$v_1$};
\node [label, below] at (1.5,2) {$v_2$};
\node [label, above] at (0.5,3.15) {$v_3$};
\end{tikzpicture}
\end{aligned}
\end{align*}

\begin{align*}
\begin{aligned}
\begin{array}{l@{\hspace{3pt}}l}
D_3.\delta &= \big[ ( v_2, [1]), (v_1, [0]), (v_3, [1]) \big]
\\[3pt]
D_3.s.\delta &= \big[(e_1, []) \big]
\end{array}
\end{aligned}
\qquad
\begin{aligned}
\begin{tikzpicture}[thick, scale=0.7]
\draw (0,0) to (0,2) to [out=right, in=down] (0.5,2.5) to (0.5,3);
\draw (0,2) to [out=left, in=down] (-0.5,2.5) to (-0.5,4);
\draw (1.5,1) to (1.5,4);
\node [blob] at (1.5,1) {};
\node [blob] at (0,2) {};
\node [blob] at (0.5,3) {};
\node [label] at (-1.2,2) {$r_1$};
\node [label] at (2.2,2) {$r_2$};
\node [label, left] at (-0.5,3) {$e_1$};
\node [label, left] at (0,0.56) {$e_1$};
\node [label, right] at (0.5,2.5) {$e_2$};
\node [label, right=-2pt] at (1.5,3) {$e_2$};
\node [label, above] at (0,2.15) {$v_1$};
\node [label, below] at (1.4,0.85) {$v_2$};
\node [label, above] at (0.5,3.15) {$v_3$};
\end{tikzpicture}
\end{aligned}
\end{align*}

\begin{align*}
\begin{aligned}
\begin{array}{l@{\hspace{3pt}}l}
D_4.\delta &= \big[ (v_1, [0]), (v_2, [1]), (v_3, [2]) \big]
\\[3pt]
D_4.s.\delta &= [(e_1, [])]
\end{array}
\end{aligned}
\qquad
\begin{aligned}
\begin{tikzpicture}[thick, scale=0.7]
\draw (0,0) to (0,1) to [out=right, in=down] (1,2) to (1,3);
\draw (0,1) to [out=left, in=down] (-1,2) to (-1,4);
\draw (0,2) to (0,4);
\node [blob] at (0,2) {};
\node [blob] at (0,1) {};
\node [blob] at (1,3) {};
\node [label] at (-1.6,2) {$r_1$};
\node [label] at (2,2) {$r_2$};
\node [label, left=-2pt] at (-1,3) {$e_1$};
\node [label, left] at (0,0.56) {$e_1$};
\node [label, right=-2pt] at (1,2.2) {$e_2$};
\node [label, left=-2pt] at (0,3) {$e_2$};
\node [label, below right] at (0,1) {$v_1$};
\node [label, below] at (0,1.85) {$v_2$};
\node [label, above] at (1,3.15) {$v_3$};
\end{tikzpicture}
\end{aligned}
\end{align*}
These 2\-diagrams consist of the same components and differ only in the order in which they have been composed. This is reflected in the different numerical values that specify the position of a node in a horizontal slice, and the different orders in which the vertices $v_1$, $v_2$ and $v_3$ appear.

Although these diagrams are isotopic, their encodings are clearly distinct, consistent with the non-strict approach we are taking. The isotopies between them arise as 3\-cell rewrites in our approach, as described in Section~\ref{sec:homotopy}.

\subsection{Algorithms}
\label{sec:algorithms}

The operation of \globular is predicated on a variety of algorithms, which we sketch here. In each case we give an indication of the type of the function which is computed, and of the procedure employed. Most algorithms run in linear time.

\paragraph{Equal.} The equality algorithm identifies whether two diagrams are identical.
$$\textbf{Equal} \big( \Delta:\Diag(n, \sigma), \Delta':\Diag(n, \sigma) \big) : \mathrm{Bool}$$
For $\Delta, \Delta'$ we first recursively compare whether $\Delta.s$ and $\Delta'.s$ are equal. If not, return \textbf{false}. Otherwise, we compare corresponding elements of $\Delta.\delta$ and $\Delta'.\delta$ sequentially. If there is a pair $\Delta.\delta[k], {\Delta'.\delta}[k]$ such that the type or coordinate data is not equal, then return \textbf{false}, otherwise return \textbf{true}. Linear time in the sizes of the diagrams.

\paragraph{Identity.} Given an $n$\-diagram, constructs the identity $(n{+}1)$-diagram.
$$\textbf{Identity} \big( \Delta:\Diag(n, \sigma)\big) : \Diag(n+1, \sigma)$$
The set of generators $\textbf{Identity}(\Delta).\delta$ is empty, and $\textbf{Identity}(\Delta).s$ is set to $\Delta$. We perform a fixed number of assignments, so the procedure terminates in constant time.

\paragraph{Rewrite.} Modifies a diagram by removing a subdiagram, and replacing it with a different subdiagram.
$$\textbf{Rewrite} (\Delta:\Diag(n, \sigma), \Psi:\Diag(n, \sigma), \Psi':\Diag(n, \sigma), C:\List(\mathbb N) ) : \Diag(n, \sigma)$$
Here $\Delta$ is the diagram that is being rewritten, $\Psi$ is the source of the rewrite, $\Psi'$ is the target of the rewrite, and $C$ is the list of coordinates specifying where the rewrite is to be applied. A total of $|\Psi.\delta|$ consecutive rewrites in $\Delta.\delta$ are removed, with the rewrites in $\Psi'.\delta$ inserted, with their coordinates offset by $C$. We illustrate this with a simple example, where $C$ is denoted by the dashed rectangle:
\[
\begin{array}{r@{{}={}}l}
\Psi
&
\begin{aligned}
\begin{tikzpicture}[thick, yscale=0.4, xscale=0.8]
\draw (0.3,1) to (0.7,2);
\draw (0.7,3) to (0.7,2);
\draw (0.3,1) to (0.3,0);
\draw (0.7,2) to [out = right, in=up](1, 1.5)to (1, 0);
\draw (0.3,1) to [out = left, in=down](0, 1.5)to (0, 3);
\node [blob] at (0.3,1) {};
\node [blob] at (0.7,2) {};
\end{tikzpicture}
\end{aligned}
\\
\Psi'
&
\begin{aligned}
\begin{tikzpicture}[thick, yscale=0.4, xscale=0.8]
\draw [white] (0, 0) to (0, 3);
\draw (0,0) to [out = up, in = left](0.5,1)
to [out = right, in = up](1,0);
\draw (0,3) to [out = down, in = left](0.5,2)
to [out = right, in = down](1,3);
\draw (0.5,1) to (0.5, 2);
\node [blob] at (0.5,1) {};
\node [blob] at (0.5,2) {};
\end{tikzpicture}
\end{aligned}
\end{array}
\hspace{1cm}
\Delta
=
\begin{aligned}
\begin{tikzpicture}[thick, yscale=0.6]
\draw (0.3,1) to (0.7,2);
\draw (0.7,3) to (0.7,2);
\draw (0.3,1) to (0.3,0);
\draw (0.7,2) to [out = right, in=up](1, 1.5)to (1, -1.5);
\draw (0.3,1) to [out = left, in=down](0, 1.5)to (0, 3);
\draw (0,-1.5) to (0,-0.5) to [out = up, in = left](0.3,0)
to [out = right, in = up](0.6,-0.5) to (0.6, -1.5);
\draw (1.5,-1.5) to (1.5,3);
\node [blob] at (0.3,1) {};
\node [blob] at (0.7,2) {};
\node [blob] at (0.3,0) {};
\node [blob] at (0.6,-1) {};
\node [rectangle, minimum width = 40pt, minimum height = 35pt, draw, fill = none, color = red, dashed] at (0.5,1.5) {};
\end{tikzpicture}
\end{aligned}
\hspace{1cm}
\textbf{Rewrite}(\Delta, \Psi, \Psi', C) =
\begin{aligned}
\begin{tikzpicture}[thick, yscale=0.6]
\draw (0.6,1) to (0.6,2);
\draw (0.9,3) to (0.9,2.5) to [out = down, in = right](0.6, 2) to [out = left, in = down](0.3, 2.5) to (0.3, 3);
\draw (0.3,0) to (0.3,0.5) to [out = up, in = left](0.6, 1) to [out = right, in = up](0.9, 0.5) to (0.9, -1.5);
\draw (0,-1.5) to (0,-0.5) to [out = up, in = left](0.3,0)
to [out = right, in = up](0.6,-0.5) to (0.6, -1.5);
\draw (1.5,-1.5) to (1.5,3);
\node [blob] at (0.6,1) {};
\node [blob] at (0.6,2) {};
\node [blob] at (0.3,0) {};
\node [blob] at (0.6,-1) {};
\end{tikzpicture}
\end{aligned}
\]
In the procedure of removing generators of $\Psi$ from $\Delta$ and inserting generators of $\Psi'$ instead, every cell is processed at most once. Numerical values in each embedding in $\Psi'$ get augmented at most once, hence the procedure is linear in the size of diagrams $\Psi$ and $\Psi'$.

\paragraph{Attach.} Attach a diagram to another diagram.
$$\textbf{Attach} (\Delta:\Diag(n, \sigma), \Delta':\Diag(k, \sigma), P:\{s,t\}, C:\List(\mathbb N)) : \Diag(n, \sigma)$$
This procedure is the implementation of the operation of diagram composition. The term `attachment' is used to indicate the effect the procedure has on the diagram in the workspace, where a visual effect of attaching a diagram is created. We attach the diagram $\Delta'$ to the diagram $\Delta$. The boolean $P$ indicates whether we are attaching to a source or the target boundary. The list $C$ describes an embedding of the source or target of $\Delta'$ in the appropriate source or target of $\Delta$, depending on $P$. 

The procedure is executed as follows:
\begin{itemize}
\item If $n-k=0$, depending on the value of $P$, we either append the elements in the lists of generators and embeddings of $\Delta'$ at the end ($P=t$) or the beginning ($P=s$) of $\Delta$'s corresponding lists. We use the numerical data in $e$ to offset the coordinates in each $\Delta[i].e$. Additionally, if $P=s$, the source boundary needs to be modified, so it is rewritten using elements  $S[i].g$ as rewriting cells.

\item If $n-k>0$, the procedure is called recursively for $\Delta.s$, with $\Delta'$, $P$ and $e$ as parameters. After the recursive call concludes, for $0\leq i\leq |D|$ we augment $\Delta[i].e$ by the offset created by adding new $(n-1)$\-cells to $\Delta.s$.
\end{itemize}
\noindent
Note that this procedure corresponds to first implicitly whiskering\footnote{In higher category theory, whiskering refers to the process of `padding' a diagram by adding identity wires at one of the sides.} $\Delta'$, so that its appropriate source or target matches that of $\Delta$, and then composing $\Delta'$ with $\Delta$ in the usual way.

In the procedure, we need to process every element in $\Delta'$ at most once, when the element gets added to the appropriate part of $\Delta$. In the scenario where $\Delta'$ is attached to the source boundary of $\Delta$, additionally the rewriting procedure needs to be performed $|\Delta'|$ times on the source of $\Delta$. It is this second step which is more costly, hence overall, the number of operations is bounded by the time complexity of performing the additional rewrites, \emph{i.e.} $|\Delta'| |\Delta.s|$.

In the example below, $\Delta'$ is the diagram being attached, $\Delta$ is the diagram we are attaching to, the boundary and the specific coordinates of the attachment point are illustrated by the blue dashed rectangle. Note that $\Delta$ and $\Delta'$ have the same dimension.
\begin{align*}
\Delta'\quad=\quad
\begin{aligned}
\begin{tikzpicture}[thick, yscale=0.4, xscale=0.8]
\draw [white] (0, 0) to (0, 3);
\draw (0,0) to [out = up, in = left](0.5,1)
to [out = right, in = up](1,0);
\draw (0,3) to [out = down, in = left](0.5,2)
to [out = right, in = down](1,3);
\draw (0.5,1) to (0.5, 2);
\node [blob] at (0.5,1) {};
\node [blob] at (0.5,2) {};
\end{tikzpicture}
\end{aligned}
&&
\Delta\quad=\quad
\begin{aligned}
\begin{tikzpicture}[thick, yscale=0.4, xscale=0.8]
\draw (0,0) to (0,5);
\draw (1,2) to (1,5);
\draw (2,0) to (2,5);
\draw  (1,5) to (1,5);
\draw  (4,5) to (4,5);
\draw (0.5,0) to (0.5,1) to [out=up, in=left](1, 2) 
to [out=right, in =up](1.5, 1) to (1.5, 0);
\draw (3,5) to (3, 4) to (3.5, 3) to  [out=right, in =down] (4, 4) to (4, 5);
\draw (3.5,0) to (3.5,3);
\draw (3,4) to [out=left, in=up](2.5,3) to (2.5, 0);
\node [blob] at (0,1) {};
\node [blob] at (1,2) {};
\node [blob] at (3,4) {};
\node [blob] at (3.5, 3) {};
\node [draw, color = blue, dashed,  minimum width=13mm, minimum height=3mm] () at (2.5,5) {};
\end{tikzpicture}
\end{aligned}
\end{align*}
The resulting diagram is as follows, where $\Delta'$ is denoted by the blue dashed rectangle.
\begin{align*}
\text{Attach}(\Delta, \Delta', t, [2])\quad=\quad
\begin{aligned}
\begin{tikzpicture}[thick, yscale=0.4, xscale=0.8]
\draw (0,0) to (0,8);
\draw (1,2) to (1,5);
\draw (2,0) to (2,5);
\draw (2.5,6) to (2.5, 7);
\draw (2,5) to [out=up, in=left] (2.5,6)
to [out=right, in =up](3, 5);
\draw (2,8) to [out=down, in=left] (2.5,7)
to [out=right, in =down](3, 8);
\draw  (1,5) to (1,8);
\draw  (4,5) to (4,8);
\draw (0.5,0) to (0.5,1) to [out=up, in=left](1, 2) 
to [out=right, in =up](1.5, 1) to (1.5, 0);
\draw (3,5) to (3, 4) to (3.5, 3) to  [out=right, in =down] (4, 4) to (4, 5);
\draw (3.5,0) to (3.5,3);
\draw (3,4) to [out=left, in=up](2.5,3) to (2.5, 0);
\node [blob] at (0,1) {};
\node [blob] at (1,2) {};
\node [blob] at (3,4) {};
\node [blob] at (3.5, 3) {};
\node [blob] at (2.5,6) {};
\node [blob] at (2.5, 7) {};
\node [draw, color = blue, dashed,  minimum width=13mm, minimum height=13mm] () at (2.5,6.5) {};
\end{tikzpicture}
\end{aligned}
\end{align*}

We can also perform attachment in the case that the diagrams $\Delta$ and $\Delta'$ have different dimensions.  Consider the following case:
\begin{align*}
\Delta'\quad&=\quad
\begin{aligned}
\begin{tikzpicture}[thick, scale = 0.7]
\draw (-1,0) to (1, 0);
\node [blob] at (0,0) {};
\node [label] at (0,-0.4) {$e_2$};
\node [label] at (0.6,0.3) {$r_2$};
\node [label] at (-0.6,0.3) {$r_2$};
\end{tikzpicture}
\end{aligned}
&
\Delta\quad&=\quad
\begin{aligned}
\begin{tikzpicture}[thick, scale=0.7]
\draw (0,0) to (0,2) to [out=right, in=down] (0.5,2.5) to (0.5,3);
\draw (0,2) to [out=left, in=down] (-0.5,2.5) to (-0.5,4);
\draw (1.5,1) to (1.5,4);
\node [blob] at (1.5,1) {};
\node [blob] at (0,2) {};
\node [blob] at (0.5,3) {};
\end{tikzpicture}
\end{aligned}
\end{align*}
Then the attachment is as follows:
\begin{calign}
\nonumber
\textbf{Attach}(\Delta, \Delta', s, [])
\quad=\quad
\begin{aligned}
\begin{tikzpicture}[thick, scale=0.7]
\draw (0,0) to (0,2) to [out=right, in=down] (0.5,2.5) to (0.5,3);
\draw (0,2) to [out=left, in=down] (-0.5,2.5) to (-0.5,4);
\draw (1.5,1) to (1.5,4);
\draw (2.5,0) to (2.5,4);
\node [blob] at (1.5,1) {};
\node [blob] at (0,2) {};
\node [blob] at (0.5,3) {};
\end{tikzpicture}
\end{aligned}
\end{calign}
Here we provide no attachment coordinates, since the attachment is to the source of $\Delta$, which is a 1\-diagram.

\paragraph{Slice.} Given an $n$-diagram, slice through it at a given height to obtain an $(n{-}1)$-diagram.
$$\textbf{Slice} (\Delta:\Diag(n, \sigma), k:\N) : \Diag(n-1, \sigma.\sigma)$$
Given an $n$\-diagram $\Delta$, we can rewrite the source boundary $\Delta.s$ using the initial $k$ entries in its list of generators $\Delta.\delta$. This gives us the $k$th \textit{slice} of $\Delta$. To execute the procedure we rewrite $\Delta.s$, using elements in $\Delta$'s lists of generators and embeddings, $k$ times. As we perform the procedure of rewriting on $(n-1)$\-diagrams, the procedure requires on the order of $k |\Delta.s|$ operations.

The \textit{source} of a diagram is its initial slice $\textbf{Slice}(\Delta,0)=\Delta.s$. The \textit{target} of a diagram is its final slice $\textbf{Slice}(\Delta, |\Delta.\delta|)$. An important note is that the resulting $(n-1)$\-diagram may be given as input to another instance of the procedure. This way, we may obtain a slice of $\Delta$ of an arbitrary dimension and location.

\paragraph{Match.} Find all the ways that one diagram appears as a subdiagram of another.
$$\textbf{Match} \big( \Delta:\Diag(n, \sigma), \Delta':\Diag(n, \sigma)) : \List(\List(\mathbb N))$$
Given two $n$\-diagrams $\Delta',\Delta$, this procedure lists all the individual instances of $\Delta'$ being a subdiagram of $\Delta$. 

First we want to find a height for the match, \textit{i.e}. an index $h$ such that $\Delta'.s$ is a subdiagram of $\textbf{Slice}(\Delta,h)$. 
For this we call the procedure recursively for $\Delta'.s$ and $\textbf{Slice}(\Delta,h)$. Given a list of such embeddings there are two possibilities. 
\begin{itemize}
\item If the list $\Delta'.\delta$ is non\-empty, we select the unique embedding consistent with the source of the generator $\Delta.\delta[h]$, let us refer to it as $e'$. We then proceed to comparing elements $\Delta'.\delta[j].a$ and $\Delta'.\delta[j].c$ with $\Delta.\delta[h+j].a$ and $\Delta.\delta[h+j].c$. If any of these checks return a mismatch, the embedding is discarded. Otherwise, an embedding of $\Delta'$ in $\Delta$ has been found and we append $h$ to the list of numerical values of $e'$ to obtain the embedding $e$. Since, we are interested in finding all embeddings of $\Delta'$ in $\Delta$, the procedure is repeated for all $0\leq h\leq |\Delta|{-}|\Delta'|$. 

\item If the list $\Delta'.\delta$ is empty, then we promote all the embeddings of $\Delta'.s$ in $\Delta[h].d$ to embeddings of $\Delta'$ in $\Delta$ by appending $h$ to the list of numerical values for each embedding.
\end{itemize}
For every recursive call, for $n$\-diagrams $\Delta', \Delta$ the procedure conducts at most $|\Delta|{-}|\Delta'|$ matching operations on diagrams, and calls itself recursively each time. In the worst case scenario, when an $n$\-diagram $\Delta'$ consists of a single 0\-cell, that results in exponential running time. However, for an $n$\-diagram $\Delta'$ whose list of generators is non\-empty, after each recursive call, we only select one match consistent with the structure of $\Delta$. This ensures that the running time is polynomial in the size of $\Delta$ and $\Delta'$.

We illustrate enumeration with the following example:
\begin{align*}
\begin{aligned}
\Delta'
\end{aligned}
&=
\begin{aligned}
\begin{tikzpicture}[thick, yscale=0.7]
\draw [white] (0, -0.5) to (0, 2.5);
\draw (0,0) to [out = up, in = left](0.5,1)
to [out = right, in = up](1,0);
\draw (0.5,1) to (0.5, 2);
\node [blob] at (0.5,1) {};
\end{tikzpicture}
\end{aligned}
&
\begin{aligned}
\Delta
\end{aligned}
&=
\begin{aligned}
\begin{tikzpicture}[thick, yscale=0.7]
\draw (0,0) to [out = up, in = left](0.5,1)
to [out = right, in = up](1,0)
to (1, -1);
\draw (0.5,1) to (0.5, 2);
\draw (-0.5,-1) to [out = up, in = left] (0, 0)
to [out = right, in = up](0.5, -1);
\node (a) [blob] at (0.5,1) {};
\node (b) [blob] at (0,0) {};
\end{tikzpicture}
\end{aligned}
&
\text{\textbf{Match}}(\Delta, \Delta')=
\begin{aligned}
\begin{tikzpicture}[thick, yscale=0.7]
\draw (0,0) to [out = up, in = left](0.5,1)
to [out = right, in = up](1,0)
to (1, -1);
\draw (0.5,1) to (0.5, 2);
\draw (-0.5,-1) to [out = up, in = left] (0, 0)
to [out = right, in = up](0.5, -1);
\node (a) [blob] at (0.5,1) {};
\node (b) [blob] at (0,0) {};
\node [rectangle, minimum width = 30pt, minimum height = 15pt, draw, fill = none, color = green, dashed] at (a) {};
\node [rectangle, minimum width = 30pt, minimum height = 15pt, draw, fill = none, color = green, dashed] at (b) {};
\end{tikzpicture}
\end{aligned}
\end{align*}
If the returned list is non-empty, we can infer that $\Delta'$ is a subdiagram of $\Delta$. The procedure of enumeration is used as pre\-processing step for rewriting and attachment, to obtain the embedding that needs to be supplied as the input for each of these procedures. If more than one option is available for the given pair of selected diagrams, the user is prompted to select the desired embedding.

As discussed above for rewriting, for a diagram $\Delta$ and a rewrite defined by $\Psi'$ and $\Psi$, enumeration looks for embeddings of $\Psi$ in $\Delta$. For attachment, for a diagram $\Delta'$ being attached to $\Delta$, enumeration looks for embeddings of $\Delta'.s$ in the appropriate target of $\Delta$, \emph{and} embeddings of $\textbf{Slice}(\Delta', |\Delta'.\delta|)$ in the appropriate source of $\Delta$. Selection of an embedding of one of these types additionally supplies the boolean indicating whether $\Delta'$ is being attached to a source or to the target of $\Delta$, which is a required input for attachment.

\subsection{Homotopies}
\label{sec:homotopy}

\def\HT{\text{HT}\xspace}

\globular also has procedures which generate homotopy moves. These are fixed families of rewrites, labelled $\I$ to $\VI$, which can be interpreted as topological diagram deformations. Within each family, several variant moves are available, which are disambiguated by subscripts. This data gives the \textit{type} of the move; we write \HT for  the set of permissible types, defined as follows:
\begin{align*}
\HT ={} &\big\{ \I_i \,|\, i \in \{1,2\}\big\}
\cup \big\{ \II_i \,|\, i \in \{1,\ldots, 8\} \big\}
\cup \big\{ \III_i \,|\, i \in \{1,\ldots, 16\} \big\}
\\ & \cup \big\{ \IV_i \,|\, i \in \{1,\ldots, 16\} \big\}
\cup \big\{ \V_i \,|\, i \in \{1,\ldots, 16\} \big\}
\cup \big\{ \VI_i \,|\, i \in \{1,\ldots, 8\} \big\}
\end{align*}
Additional complexity arises from the fact that, even after specifying the move type, the rewrite it gives rise to is \textit{contextual}, meaning that it depends on the geometry of the diagram to which it is being applied, and to the chosen location within the diagram, given as a list of coordinates.

To fully specify a homotopy move, one must therefore specify a diagram, a list of coordinates, and a move type. In \globular, this data is used as input for the following functions.
\begin{itemize}
\item $\textbf{HomotopyMatch}(\Delta: \Diag(n,\sigma), p:\N^*, t:\HT) : \text{Boolean}$

\noindent Returns true if a homotopy move of type $t$ is admissible at position $p$ in diagram $\Delta$, and false otherwise.

\item $\textbf{HomotopyBox}(\Delta: \Diag(n,\sigma), h:\N^*, t:\HT) : \text{Pair(\N, \N)}$

\noindent
If \textbf{HomotopyMatch} returns true on this data, this function returns the first and last positions in the list of elements $\Delta.\delta$ affected by this homotopy. Otherwise, the behaviour is undefined.

\item $\textbf{HomotopyRewrite}(\Delta: \Diag(n,\sigma), h:\N^*, t:\HT) : \Diag(n,\sigma)$

\noindent
If \textbf{HomotopyMatch} returns true on this data, this function returns the diagram resulting from acting on $\Delta$ by the indicated homotopy move. Otherwise, the behaviour is undefined.
\end{itemize}
In \globular, these functions are used to extend as appropriate the algorithms presented in Section~\ref{sec:algorithms}. For example, the \textbf{HomotopyRewrite} function may be used in place of the \textbf{Rewrite} function, when the move to be applied is a homotopy move, rather than a move arising from the signature.

We illustrate here the form of the Type {\I} and Type {\II} moves. The form for the other types is more involved, and we refer to the paper~\cite{globular-theory} for further details.

\subsection*{Type I} As rewrites, these homotopy moves exchange the heights of non-interacting vertices:
\begin{equation}
\label{eq:typeIexample}
\begin{aligned}
\begin{tikzpicture}[thick, yscale=0.8]
\draw [region] (-0.5,0.2) rectangle +(2,2.55);
\draw (0,0.2) to (0,2.75);
\draw (1,0.2) to (1,2.75);
\node [morphism] at (0,2) {$f$};
\node [morphism] at (1,1) {$g$};
\end{tikzpicture}
\end{aligned}
\quad\begin{array}{@{}c@{}}\I_1\\\rightarrow\\\leftarrow\\\I_2\end{array}\quad
\begin{aligned}
\begin{tikzpicture}[thick, yscale=0.8]
\draw [region] (-0.5,0.2) rectangle +(2,2.55);
\draw (0,0.2) to (0,2.75);
\draw (1,0.2) to (1,2.75);
\node [morphism] at (0,1) {$f$};
\node [morphism] at (1,2) {$g$};
\end{tikzpicture}
\end{aligned}
\end{equation}
In general, we allow $f$ and $g$ to have an arbitrary number of input and output wires, and we allow any number of wires between $f$ and $g$.

As elements of diagrams themselves, they are drawn in the following graphical style:
\begin{calign}
\nonumber
\begin{aligned}
\begin{tikzpicture}[thick, scale=0.8]
\draw [region] (-0.2,0) rectangle +(2.4,2);
\draw (0.6,0) to [out=up, in=down] (1.6,1.8) to (1.6,2);
\draw [region] (0.0,-0.2) rectangle +(2.4,2);
\draw (1.6,-0.2) to (1.6,0) to [out=up, in=down] (0.6,1.8);
\end{tikzpicture}
\end{aligned}
&
\begin{aligned}
\begin{tikzpicture}[thick, scale=0.8]
\draw [region] (-0.2,0) rectangle +(2.4,2);
\draw (1.6,0) to [out=up, in=down] (0.6,1.8) to (0.6,2);
\draw [region] (0.0,-0.2) rectangle +(2.4,2);
\draw (0.6,-0.2) to (0.6,0) to [out=up, in=down] (1.6,1.8);
\end{tikzpicture}
\end{aligned}
\\\nonumber
\I_1 & \I_2
\end{calign}
The user-interface command to implement a Type I homotopy move is clicking-and-dragging one of the vertices up or down. For example, to trigger move $\I_1$ as illustrated in expression~\ref{eq:typeIexample}, then given the left-hand diagram, the user could drag the $f$ vertex down, or the $g$ vertex up.

\subsection*{Type II}

These describe naturality of Type \I\ moves:
\begin{center}
$
\begin{aligned}
\begin{tikzpicture}[thick, scale=0.8]
\draw [region] (-0.2,-1) rectangle +(2.4,3);
\draw (0.6,-1) to (0.6,0) to [out=up, in=down] (1.6,2);
\node [morphism, anchor=north] at (0.6,0) {$\alpha$};
\begin{scope}[]
\draw [region] (0,-1.2) rectangle +(2.4,3);
\end{scope}
\draw (1.6,-1.2) to (1.6,0) to [out=up, in=down] (0.6,1.8);
\end{tikzpicture}
\end{aligned}
\quad\begin{array}{@{}c@{}}\II_1\\\rightarrow\\\leftarrow\\\II_2\end{array}\quad
\begin{aligned}
\begin{tikzpicture}[thick, scale=0.8]
\draw [region] (-0.2,0) rectangle +(2.4,3);
\draw (0.6,0) to [out=up, in=down] (1.6,1.8) to (1.6,3);
\node [morphism, anchor=south] at (1.6,1.8) {$\alpha$};
\draw [region] (0.0,-0.2) rectangle +(2.4,3);
\draw (1.6,-0.2) to (1.6,0) to [out=up, in=down] (0.6,1.8) to (0.6,2.8);
\end{tikzpicture}
\end{aligned}
\hspace{1cm}
\begin{aligned}
\begin{tikzpicture}[thick, scale=0.8]
\draw [region] (-0.2,-1) rectangle +(2.4,3);
\draw (0.6,-1) to (0.6,0) to [out=up, in=down] (1.6,2);
\begin{scope}[]
\draw [region] (0,-1.2) rectangle +(2.4,3);
\end{scope}
\draw (1.6,-1.2) to (1.6,0) node [morphism, anchor=north] {$\alpha$} to [out=up, in=down] (0.6,1.8);
\end{tikzpicture}
\end{aligned}
\quad\begin{array}{@{}c@{}}\II_3\\\rightarrow\\\leftarrow\\\II_4\end{array}\quad
\begin{aligned}
\begin{tikzpicture}[thick, scale=0.8]
\draw [region] (-0.2,0) rectangle +(2.4,3);
\draw (0.6,0) to [out=up, in=down] (1.6,2) to (1.6,3);
\draw [region] (0.0,-0.2) rectangle +(2.4,3);
\draw (1.6,-0.2) to (1.6,0) to [out=up, in=down] (0.6,1.8) to (0.6,2.8);
\node [morphism, anchor=south] at (0.6,1.8) {$\alpha$};
\end{tikzpicture}
\end{aligned}
$
\end{center}
In general we allow $\alpha$ to have any number of input and output wires, and we allow for any number of interleaving sheets. Note that the source and target diagrams here themselves make use of Type \I\ moves. The user interface command is clicking-and-dragging the appropriate vertices labelled $\alpha$  above, making them intuitive to execute.

\section{Technology}
\label{sec:technology}

Here we describe technological aspects of the implementation.
\globular is implemented in Javascript and runs client-side embedded in the web browser, with all the computation taking place on the user's machine, therefore limiting the need for data transfer. The back-end is a \texttt{Node.js} server, responding to user requests and hosting an account system that allows users to register and privately save working versions of proofs, allowing work to be continued on a different machine. However, there is no requirement to register for an account to use the tool.

The tool can encode the current signature in Javascript Object Notation (JSON), allowing export as a plain text file, compressed in LZ4 format to reduce the file size. Along with the corresponding import operation, this allows users to back-up their work on their local machine.  When the user is satisfied with the finished proof, they can make it public and share it with the rest of community; the proof is then added to the \globular public gallery, and a unique URL linking to the proof is generated. We give examples of this functionality at the end of this chapter. The entire project is open-source, and the code is available at \url{globular.science/source}.

While Javascript has a weak type-theoretic structure, formal verification of the code is not a priority for us. The present implementation should be seen as a proof-of-concept prototype, that the theoretical basis outlined in~\cite{globular-theory}. We hope that future iterations of the tool will be more amenable to formal verification. Our main immediate goal has been to produce a tool which is useful for the community, and in that it seems we have been reasonably successful.

The interface has been designed to be friendly and intuitive. Diagrams can be created, rewritten and composed by clicking elements in the signature and selecting an attachment point from the list of options. Given an $n$\-diagram $D$ in the workspace, the operation triggered depends on the dimension $k$ of the cell $g$ that we select from the signature $\Sigma$. If $k=n+1$, then $D$ is rewritten; if $k<n+1$, then $g$ is attached to $D$. For the latter, first implicitly a diagram $S=i(g)$ of the generator $g$ is created. Diagrams $S$ and $D$ then get composed in accordance with the \textbf{Attach} procedure described above. If $k>n+1$ there is no effect on $D$ and the tool asks the user to select another cell.

However, selecting elements from the signature is not the only method of modifying the diagram in the signature. Interchanger morphisms of types $\I$-$\VI$ can be applied directly by clicking and dragging the appropriate cells within the diagram. 

The graphical visualisations of cells are generated using the vector graphics technology SVG, which is widely supported by modern browsers. However, this limits the rendering to 2 dimensions. This may be regarded as a serious difficulty, especially when dealing with higher dimensional structures. For that reason, in the future, we intend to implement a 3D graphics engine using Three.js. However, even these enhanced graphical capabilities will not be sufficient to work efficiently with structures of dimension $n=4$ and higher. To work around that, we implemented a system of toggles, that allows to suppress the lowest dimensions and view slices that are of interest. Even though, at times, this may prove cumbersome, it is certainly worthwhile as this solution provides us with a systematic method for viewing morphisms in any $n$\-dimensional structure. For an $n$\-diagram $D$ such that $n\geq 3$, for which the number of dimensions projected out is $k$, there are $n-k-2$ slice toggles that allow us to view a multidimensional structure as a sequence of 2D slices. 

\section{Examples}\label{sec:ex}

Here we give examples of formalized proofs from algebra and topology. In each case we briefly describe the mathematical context of the proof, and give some details of its formalization. Direct hyperlinks are provided to the formalized proofs on the \globular website; to navigate these proofs, use the \emph{Project} and \emph{Slice} controls at the top-right, and move your mouse cursor over the different parts of the main diagram to understand its components. Documentation on how to use \globular is available~\cite{globular}. To our knowledge, none of these results have previously been formalized by any existing tool.

\begin{example}[Frobenius implies associative, \hrefLMCS{http://globular.science/1512.004}{globular.science/1512.004}, length 12]
\textit{In a monoidal category, if multiplication and comultiplication morphisms are unital, counital and Frobenius, then they are associative and coassociative.} We formalize this in \globular using a 2\-category with a single 0\-cell, since this is algebraically equivalent to a monoidal category. Such a proof would be traditionally written out as a series of pictures; for example, see the textbook~\cite{Kock_2003}. \globular produces these pictures automatically.
\end{example}

\begin{example}[Strengthening an equivalence, \hrefLMCS{http://globular.science/1512.007}{globular.science/1512.007}, length 14]
\textit{In a 2\-category, an equivalence gives rise to an adjoint equivalence.} This is a classic result from the category theory community~\cite{baezlauda, Rivano}; it can be considered one of the first nontrivial theorems of 2\-category theory. We investigate it in further detail in Section~\ref{sec:using}.
\end{example}

\begin{example}[Swallowtail comes for free, \hrefLMCS{http://globular.science/1512.006}{globular.science/1512.006}, length 12]
\textit{In a monoidal 2-category, a weakly-dual pair of objects gives rise to a strongly-dual pair, satisfying the swallowtail equations.} This theorem plays an important role in the singularity theory of 3\-manifolds~\cite{piotr-thesis}. For the formalization, we model a monoidal 2\-category as a 3\-category with one 0\-cell.
\end{example}

\begin{example}[Pentagon and triangle implies $\rho _I = \lambda _I$, \hrefLMCS{http://globular.science/1512.002}{globular.science/1512.002}, length 62]
\textit{In a monoidal 2\-category, a pseudomonoid object satisfies $\rho _I = \lambda _I$.} A \textit{pseudomonoid} is a higher algebraic structure categorifying the concept of monoid; it has the property that a pseudomonoid in {\bf Cat} is the same as a monoidal category. Such a structure is known to be coherent~\cite{Lack_2000}, in the sense that all equations commute, and here we give an explicit proof of the equation $\rho_I = \lambda_I$, which played an important role in the early study of coherence for monoidal categories.
\end{example}

\begin{example}[The antipode is an algebra homomorphism, \hrefLMCS{http://globular.science/1512.011}{globular.science/1512.011}, length~68]
\textit{For a Hopf algebra structure in a braided monoidal category, the antipode is an algebra homomorphism.} Hopf algebras are algebraic structures which play an important role in representation theory and physics~\cite{Majid_2002, Street_2007}. Proofs involving these structures are usually presented in \textit{Sweedler notation}, a linear syntax which represents coalgebraic structures using strings of formal variables with subscripts; we do not know of any existing approaches to formal verification for Sweedler proofs. This formalization in \globular is translated from a Sweedler proof given in~\cite{pareigis}. For the formalization, we model a braided monoidal category as a 3\-category with one 0\-cell and one 1\-cell. This formalization is due to Dominic Verdon.
\end{example}

\begin{example}[The Perko knots are isotopic, \hrefLMCS{http://globular.science/1512.012}{globular.science/1512.012}, length 251]
\textit{The Perko knots are isotopic.}
The Perko knots are a pair of 10-crossing knots stated by Little in 1899 to be distinct, but proven by Perko in 1974 to be isotopic~\cite{Perko_1974}. Here we give the isotopy proof, adapted from~\cite{mathforum}. A nice feature is that the second and third Reidemeister moves do not have to be entered, since they are already implied by the 3-category axioms. The proof consists of a series of 251 atomic deformations, which rewrite the first Perko knot into the second. By stepping through the proof one rewrite at a time, the isotopy itself can be visualized as a movie.
\end{example}

\begin{example}
[Constructing the codensity monad, \hrefLMCS{http://globular.science/1611.003v2}{globular.science/1611.003v2}]
This project demonstrates the potential utility of \globular in
classical, 1-categorical applications. It reproduces the graphical
language used by Hinze~\cite{Hinze_2012} for depicting
Kan extensions by augmenting the usual 2-categorical string diagram
language of Cat with certain brackets, which indicate a
2-categorical version of `currying', sending natural
transformations $\alpha : F\circ J \to G$ to $[\alpha] : F \to
\textrm{Ran}_J(G)$. In this example we reproduce these brackets with the help of
a dummy 1-cell, and prove a standard fact about Kan extensions: the Kan extension of a functor $J$ over itself always has a monad
structure, called the \textit{codensity monad}.
\end{example}


\raggedright
\bibliographystyle{plainurl}
\bibliography{references}

\end{document}

%% file: assoc3D.tikz
\begin{tikzpicture}[dotpic, xscale=0.9, {red dot/.style}={dot,fill=red,xscale=0.8}]
	\begin{pgfonlayer}{nodelayer}
		\node [style=none] (0) at (-8, 4.75) {};
		\node [style=none] (1) at (-4.75, 1.5) {};
		\node [style=none] (2) at (-4.75, -5.75) {};
		\node [style=none] (3) at (-8, -2.5) {};
		\node [style=red dot] (4) at (-7, 2.5) {};
		\node [style=none] (5) at (-7, 3.75) {};
		\node [style=none] (6) at (-7.5, 2.25) {};
		\node [style=none] (7) at (-7.5, -3) {};
		\node [style=none] (8) at (-5.25, -0.75) {};
		\node [style=none] (9) at (-7.25, -2) {};
		\node [style=none] (10) at (-7.25, 1.25) {};
		\node [style=none] (11) at (-5.25, -4.25) {};
		\node [style=none] (12) at (-6.5, 1.25) {};
		\node [style=red dot] (13) at (-6.5, 0) {};
		\node [style=none] (14) at (-7, -0.25) {};
		\node [style=none] (15) at (-6, -1.25) {};
		\node [style=none] (16) at (-6.5, 0.75) {};
		\node [style=red dot] (17) at (-6, -2.5) {};
		\node [style=none] (18) at (-6.5, -2.75) {};
		\node [style=none] (19) at (-5.5, -3.75) {};
		\node [style=none] (20) at (-6, -1.75) {};
		\node [style=none] (21) at (-5.5, -5) {};
		\node [style=none] (22) at (-6.5, -4) {};
		\node [style=none] (23) at (-7, -3.5) {};
		\node [style=none] (24) at (1.25, 3.5) {};
		\node [style=none] (25) at (1.5, -1.25) {};
		\node [style=none] (26) at (2, 0.5) {};
		\node [style=none] (27) at (1, -2.25) {};
		\node [style=none] (28) at (3, -0.75) {};
		\node [style=none] (29) at (1, -3.5) {};
		\node [style=none] (30) at (0.75, -2) {};
		\node [style=none] (31) at (0, -2.5) {};
		\node [style=red dot] (32) at (1.25, 2.25) {};
		\node [style=none] (33) at (0, 4.75) {};
		\node [style=none] (34) at (3, -4.25) {};
		\node [style=none] (35) at (2, -3.25) {};
		\node [style=none] (36) at (1.5, -0.75) {};
		\node [style=none] (37) at (2, -4.5) {};
		\node [style=none] (38) at (0.75, 1.25) {};
		\node [style=none] (39) at (3.25, 1.5) {};
		\node [style=none] (40) at (2, 0.25) {};
		\node [style=none] (41) at (3.25, -5.75) {};
		\node [style=none] (42) at (0.5, 1.75) {};
		\node [style=none] (43) at (0.5, 2) {};
		\node [style=red dot] (44) at (2, -0.5) {};
		\node [style=red dot] (45) at (1.5, -2) {};
		\node [style=none] (46) at (2.5, -1.75) {};
		\node [style=none] (47) at (2.5, -5) {};
		\node [style=none] (48) at (0.5, -3) {};
		\node [style=red dot] (49) at (9.5, -2) {};
		\node [style=none] (50) at (8, 4.75) {};
		\node [style=none] (51) at (8.25, 3.75) {};
		\node [style=none] (52) at (9.75, 3) {};
		\node [style=none] (53) at (9.5, -1) {};
		\node [style=none] (54) at (9.5, -1.25) {};
		\node [style=none] (55) at (11.25, -5.75) {};
		\node [style=none] (56) at (8.5, 0) {};
		\node [style=none] (57) at (10, -4.5) {};
		\node [style=none] (58) at (8, -2.5) {};
		\node [style=red dot] (59) at (9.75, 1.75) {};
		\node [style=none] (60) at (9, -3.5) {};
		\node [style=none] (61) at (11.25, 1.5) {};
		\node [style=none] (62) at (9, 1) {};
		\node [style=none] (63) at (10.5, 0) {};
		\node [style=none] (64) at (8.5, -1) {};
		\node [style=none] (65) at (9, 1.5) {};
		\node [style=none] (66) at (9, -2.25) {};
		\node [style=none] (67) at (10.5, -5) {};
		\node [style=none] (68) at (10, -3.25) {};
		\node [style=none] (69) at (10.75, 1.25) {};
		\node [style=none] (70) at (10.75, -2.5) {};
		\node [style=red dot] (71) at (9, 0.25) {};
		\node [style=none] (72) at (8.5, -3) {};
		\node [style=none] (73) at (8.25, 0) {};
		\node [style=none] (74) at (0.25, 0) {};
		\node [style=none] (75) at (0.25, 3.75) {};
		\node [style=none] (76) at (2.75, -2.5) {};
		\node [style=none] (77) at (2.75, 1.25) {};
		\node [style=none] (78) at (-1.25, -2.5) {\color{blue}\footnotesize\sf assoc};
		\node [style=none] (79) at (6.75, -0.5) {\color{blue}\footnotesize\sf assoc};
		\node [style=none] (80) at (18, -1.5) {};
		\node [style=none] (81) at (16, -2.5) {};
		\node [style=none] (82) at (16, 4.75) {};
		\node [style=none] (83) at (18.25, 2.5) {};
		\node [style=red dot] (84) at (17.5, -0.25) {};
		\node [style=none] (85) at (18, -4.5) {};
		\node [style=red dot] (86) at (18.25, 1.25) {};
		\node [style=none] (87) at (17.5, 0.5) {};
		\node [style=none] (88) at (19.25, 1.5) {};
		\node [style=none] (89) at (17.5, 1) {};
		\node [style=none] (90) at (18.75, -5.25) {};
		\node [style=none] (91) at (16.5, -1.75) {};
		\node [style=none] (92) at (19.25, -5.75) {};
		\node [style=none] (93) at (16.5, -3) {};
		\node [style=red dot] (94) at (17, -1.5) {};
		\node [style=none] (95) at (17.5, -2.75) {};
		\node [style=none] (96) at (18, -2.5) {};
		\node [style=none] (97) at (17.5, -4) {};
		\node [style=none] (98) at (17, -1) {};
		\node [style=none] (99) at (18.75, -0.25) {};
		\node [style=none] (100) at (17, -0.5) {};
		\node [style=none] (101) at (8.25, 1.75) {};
		\node [style=none] (102) at (16.25, -1.75) {};
		\node [style=none] (103) at (18.5, -4) {};
		\node [style=none] (104) at (8.25, -1.75) {};
		\node [style=none] (105) at (14.25, -2.25) {\color{blue}\footnotesize\sf assoc};
		\node [style=none] (106) at (10.25, -0.5) {};
		\node [style=none] (107) at (10.25, -4) {};
		\node [style=none] (108) at (18.5, -0.5) {};
		\node [style=none] (109) at (16.25, 1.75) {};
	\end{pgfonlayer}
	\begin{pgfonlayer}{edgelayer}
		\draw [thick, style=blue] (106.center) to (101.center);
		\draw [style=gray edge] (3.center) to (2.center);
		\draw [style=gray edge] (2.center) to (1.center);
		\draw [style=gray edge] (1.center) to (0.center);
		\draw [style=gray edge] (0.center) to (3.center);
		\draw [thick, thick] (4) to (5.center);
		\draw [thick, in=165, out=90, looseness=1.25] (6.center) to (4);
		\draw [thick] (7.center) to (6.center);
		\draw [thick, style=blue] (9.center) to (11.center);
		\draw [thick, style=blue] (11.center) to (8.center);
		\draw [thick, style=blue] (8.center) to (10.center);
		\draw [thick, style=blue] (10.center) to (9.center);
		\draw [thick, in=-45, out=90, looseness=1.00] (12.center) to (4);
		\draw [thick] (13) to (16.center);
		\draw [thick, in=165, out=90, looseness=1.25] (14.center) to (13);
		\draw [thick, in=-45, out=90, looseness=1.00] (15.center) to (13);
		\draw [thick] (17) to (20.center);
		\draw [thick, in=165, out=90, looseness=1.25] (18.center) to (17);
		\draw [thick, in=-45, out=90, looseness=1.00] (19.center) to (17);
		\draw [thick] (20.center) to (15.center);
		\draw [thick] (22.center) to (18.center);
		\draw [thick] (21.center) to (19.center);
		\draw [thick] (16.center) to (12.center);
		\draw [thick] (23.center) to (14.center);
		\draw [style=gray edge] (31.center) to (41.center);
		\draw [style=gray edge] (41.center) to (39.center);
		\draw [style=gray edge] (39.center) to (33.center);
		\draw [style=gray edge] (33.center) to (31.center);
		\draw [thick] (32) to (24.center);
		\draw [thick, in=165, out=90, looseness=1.25] (43.center) to (32);
		\draw [thick, in=-90, out=90, looseness=1.00] (42.center) to (43.center);
		\draw [thick, style={blue!30}] (30.center) to (34.center);
		\draw [thick, style=blue] (34.center) to (28.center);
		\draw [thick, style=blue] (28.center) to (38.center);
		\draw [thick, style={blue!30}] (38.center) to (30.center);
		\draw [thick, in=-45, out=90, looseness=1.00] (26.center) to (32);
		\draw [thick] (44) to (40.center);
		\draw [thick, in=165, out=90, looseness=1.25] (36.center) to (44);
		\draw [thick, in=-45, out=90, looseness=1.00] (46.center) to (44);
		\draw [thick] (45) to (25.center);
		\draw [thick, in=165, out=90, looseness=1.25] (27.center) to (45);
		\draw [thick, in=-45, out=90, looseness=1.00] (35.center) to (45);
		\draw [thick] (29.center) to (27.center);
		\draw [thick] (37.center) to (35.center);
		\draw [thick, in=-90, out=90, looseness=1.00] (40.center) to (26.center);
		\draw [thick] (47.center) to (46.center);
		\draw [thick] (25.center) to (36.center);
		\draw [thick] (48.center) to (42.center);
		\draw [style=gray edge] (58.center) to (55.center);
		\draw [style=gray edge] (55.center) to (61.center);
		\draw [style=gray edge] (61.center) to (50.center);
		\draw [style=gray edge] (50.center) to (58.center);
		\draw [thick] (59) to (52.center);
		\draw [thick, in=165, out=90, looseness=1.25] (65.center) to (59);
		\draw [thick, style=blue] (70.center) to (69.center);
		\draw [thick, style=blue] (69.center) to (51.center);
		\draw [thick, style={blue!30}] (51.center) to (73.center);
		\draw [thick, in=-45, out=90, looseness=1.00] (63.center) to (59);
		\draw [thick] (71) to (62.center);
		\draw [thick, in=165, out=90, looseness=1.25] (56.center) to (71);
		\draw [thick, in=-45, out=90, looseness=1.00] (53.center) to (71);
		\draw [thick, in=165, out=90, looseness=1.25] (66.center) to (49);
		\draw [thick, in=-45, out=90, looseness=1.00] (68.center) to (49);
		\draw [thick] (60.center) to (66.center);
		\draw [thick] (67.center) to (63.center);
		\draw [thick] (64.center) to (56.center);
		\draw [thick] (62.center) to (65.center);
		\draw [thick, style=blue] (74.center) to (76.center);
		\draw [thick, style=blue] (76.center) to (77.center);
		\draw [thick, style=blue] (77.center) to (75.center);
		\draw [thick, style=blue] (75.center) to (74.center);
		\draw [thick, style=blue] (77.center) to (69.center);
		\draw [thick, style=blue] (75.center) to (51.center);
		\draw [thick, style={blue!30}] (74.center) to (73.center);
		\draw [thick, style=blue] (8.center) to (28.center);
		\draw [thick, style=blue] (11.center) to (34.center);
		\draw [thick, style=blue] (10.center) to (38.center);
		\draw [thick, style={blue!30}] (9.center) to (30.center);
		\draw [style=gray edge] (81.center) to (92.center);
		\draw [style=gray edge] (92.center) to (88.center);
		\draw [style=gray edge] (88.center) to (82.center);
		\draw [style=gray edge] (82.center) to (81.center);
		\draw [thick] (86) to (83.center);
		\draw [thick, in=165, out=90, looseness=1.25] (89.center) to (86);
		\draw [thick, in=-45, out=90, looseness=1.00] (99.center) to (86);
		\draw [thick] (84) to (87.center);
		\draw [thick, in=165, out=90, looseness=1.25] (100.center) to (84);
		\draw [thick, in=-45, out=90, looseness=1.00] (80.center) to (84);
		\draw [thick] (94) to (98.center);
		\draw [thick, in=165, out=90, looseness=1.25] (91.center) to (94);
		\draw [thick, in=-45, out=90, looseness=1.00] (95.center) to (94);
		\draw [thick] (93.center) to (91.center);
		\draw [thick] (97.center) to (95.center);
		\draw [thick] (85.center) to (96.center);
		\draw [thick] (90.center) to (99.center);
		\draw [thick] (96.center) to (80.center);
		\draw [thick] (98.center) to (100.center);
		\draw [thick] (87.center) to (89.center);
		\draw [thick, style=blue] (104.center) to (107.center);
		\draw [thick, style=blue] (107.center) to (106.center);
		\draw [thick, style=blue] (101.center) to (104.center);
		\draw [thick, style={blue!30}] (102.center) to (103.center);
		\draw [thick, style=blue] (103.center) to (108.center);
		\draw [thick, style=blue] (108.center) to (109.center);
		\draw [thick, style={blue!30}] (109.center) to (102.center);
		\draw [thick, style=blue] (106.center) to (108.center);
		\draw [thick, style=blue] (107.center) to (103.center);
		\draw [thick, style=blue] (101.center) to (109.center);
		\draw [thick, style={blue!30}] (104.center) to (102.center);
		\draw [style=gray edge] (0.center) to (82.center);
		\draw [style=gray edge] (1.center) to (88.center);
		\draw [style=gray edge] (2.center) to (92.center);
		\draw [thick] (54.center) to (53.center);
		\draw [thick] (49) to (54.center);
		\draw [thick] (72.center) to (64.center);
		\draw [thick] (57.center) to (68.center);
		\draw [thick, style={blue!30}] (73.center) to (70.center);
		\draw [thick, style=blue] (76.center) to (70.center);
	\end{pgfonlayer}
\end{tikzpicture}